\documentclass[aap,MSNbibl,seceqn,citesort,dvips]{arximspdf}
\usepackage{subenv}

\doi{10.1214/11-AAP801}
\volume{22}
\issue{4}
\pubyear{2012}
\firstpage{1541}
\lastpage{1575}

\makeatletter

\newtheorem{theorem}{Theorem}[section]
\newtheorem{lemma}{Lemma}[section]
\newproclaim{condition}{Condition}[section]
\newproclaim{assmpn}{Assumption}[section]
\newproclaim{rem}{Remark}[section]
\newproclaim{definition}{Definition}[section]

\newtheorem{cor}{Corollary}[section]

\newcommand{\eps}{{\varepsilon}}
\newcommand{\R}{{\mathbb{R}}}

\newcommand{\ep}{\varepsilon}
\renewcommand{\d}{\delta}
\newcommand{\Xedt}{X_{\ep,\d,t}}
\newcommand{\Xeds}{X_{\ep,\d,s}}
\newcommand{\Yeds}{Y_{\ep,\d,s}}
\newcommand{\Aed}{A_{\ep,\d}}
\newcommand{\ued}{u_{\ep,\d}}
\newcommand{\Hed}{H_{\ep,\d}}
\renewcommand{\H}{H_\ep}
\renewcommand{\u}{u_{\ep}}
\renewcommand{\L}{\Lambda}

\renewcommand{\bar}{\overline}

\makeatother

\begin{document}
\begin{frontmatter}

\title{Small-time asymptotics for fast mean-reverting stochastic volatility models\thanksref{T1}}
\runtitle{Asymptotics for stochastic volatility models}

\thankstext{T1}{Supported in part by NSF Grants DMS-08-06434 and
DMS-08-06461.}

\begin{aug}
\author[A]{\fnms{Jin} \snm{Feng}\thanksref{T2}\ead[label=e1]{jfeng@math.ku.edu}\ead[label=u1,url]{http://www.math.ku.edu/\textasciitilde jfeng/}},
\author[B]{\fnms{Jean-Pierre} \snm{Fouque}\ead[label=e2]{fouque@pstat.ucsb.edu}\ead[label=u2,url]{http://www.pstat.ucsb.edu/faculty/fouque/}}
\and
\author[B]{\fnms{Rohini} \snm{Kumar}\corref{}\ead[label=e3]{kumar@pstat.ucsb.edu}\ead[label=u3,url]{http://www.pstat.ucsb.edu/faculty/kumar/}}
\runauthor{J. Feng, J.-P. Fouque and R. Kumar}
\affiliation{University of Kansas, University of California, Santa
Barbara, and~University of California, Santa Barbara}
\address[A]{J. Feng\\
Department of Mathematics\\
University of Kansas\\
Lawrence, Kansas 66045\\
USA\\
\printead{e1}\\
\printead{u1}}
\address[B]{J.-P. Fouque \\
R. Kumar\\
Department of Statistics\\
\quad and Applied Probability\\
University of California\\
Santa Barbara, California 93106\\
USA\\
\printead{e2}\\
\hphantom{E-mail: }\printead*{e3}\\
\printead{u2}\\
\hphantom{URL: }\printead*{u3}} 
\end{aug}

\thankstext{T2}{Supported by funding from the State of Kansas
Kan0064212.}

\received{\smonth{9} \syear{2010}}
\revised{\smonth{4} \syear{2011}}

%
\begin{abstract}
In this paper, we study stochastic volatility models in regimes where
the maturity is small, but large compared to the mean-reversion time of
the stochastic volatility factor. The problem falls in the class of
averaging/homogenization problems for nonlinear HJB-type equations
where the ``fast variable'' lives in a noncompact space. We develop a
general argument based on viscosity solutions which we apply to the two
regimes studied in the paper. We derive a large deviation principle,
and we deduce asymptotic prices for out-of-the-money call and put
options, and their corresponding implied volatilities. The results of
this paper generalize the ones obtained in Feng, Forde and Fouque
[\textit{SIAM J. Financial Math.} \textbf{1} (2010) 126--141] by a
moment generating function computation in the particular case of the
Heston model.
\end{abstract}

%

\end{frontmatter}

\section{Introduction}\label{intro}

On one hand, the theory of large deviations has been recently applied
to local and stochastic volatility models
\cite{Avell02,Avell03,BBF-LV,BBF-SV,HL05} and has given very
interesting results on the behavior of implied volatilities near
maturity. (An implied volatility is the volatility parameter needed in
the Black--Scholes formula in order to match a call option price; it is
common practice to quote prices in volatility through this
transformation.) In the context of stochastic volatility models, the
rate function involved in the large deviation estimates is given in
terms of a distance function, which in general cannot be calculated in
closed form. For particular models, such as the SABR model
\cite{Hagan,HL-Risk07}, approximations obtained by expansion techniques
have been proposed; see also \cite{gatheral,HL-book,lewis}.
Semi closed form expressions for short time implied volatilities have
been obtained in \cite{FJ09}.

On the other hand, multi-factor stochastic volatility models have
been studied during the last ten years by many authors (see, e.g.,
\cite{duffie,FPS00,gatheral,lebaron,Bouchaud}). They
are quite efficient in capturing the main features of implied
volatilities known as smiles and skews, but they are usually not
simple to calibrate. In the presence of separated time scales, an
asymptotic theory has been proposed in \cite{FPS00,FPSS03}. It has
the advantage of capturing the main effects of stochastic volatility
through a small number of group parameters arising in the asymptotic.
The fast time scale expansion is related to the ergodic property of the
corresponding fast mean-reverting stochastic volatility factor.

It is natural to try to combine these two modeling aspects and limiting
results, by considering short maturity options computed with fast
mean-reverting stochastic volatility models, in such a way that
maturity is of order $\eps\ll1$, and the mean-reversion time,
$\delta$, of volatility is even smaller of order $\delta=\eps^2$ (fast
mean-reversion) or $\delta=\eps^4$ (ultra-fast mean-reversion).

In \cite{FFF}, the authors studied the particular case of the Heston
model in the regime $\delta=\eps^2$ by an explicit computation of the
moment generating function of the stock price and its asymptotic
analysis.

In this paper, we establish a large deviation principle for general
stochastic volatility models in the two regimes of fast and ulta-fast
mean-reversion, and we derive asymptotic smiles/skews. For such general
dynamics, a moment generating function approach is no longer available.
Our problem falls in the class of homogenization/averaging problems for
nonlinear HJB-type equations where the ``fast variable'' lives in a
noncompact space. We develop a general argument based on viscosity
solutions which we apply to the two regimes studied in the paper.
Viscosity solution techniques have been used in averaging of nonlinear
HJB equations over noncompact space in \cite{BCM10}. However, the
techniques in \cite{BCM10} were proved for a certain class of
nonlinear HJB equations which does not include our case. In this
paper, we develop a method more general than \cite{BCM10}. In
particular, it can be used to treat the problems in \cite{BBF-SV}, but
not vice versa.

We start by considering the following stochastic differential equations
modeling the evolution of the stock price $(S_t)$ under a risk-neutral
pricing probability measure, and with a stochastic volatility
determined by a~process~$(Y_t)$:
%
%
\begin{subequation}\label{sde-1}
%
%
\begin{eqnarray}
\label{S1}
dS_t &=& r S_t\,dt+\sigma(Y_t )S_t \,dW_t^{(1)},\\
\label{Y1}
dY_t &=& \frac{1}{\delta}(m-Y_t) \,dt + \frac{\nu}{\sqrt{\delta}}
Y^\beta_t d W^{(2)}_t,
\end{eqnarray}
\end{subequation}
where $m\in\mathbb R, r,\nu>0$, $W^{(1)}$ and $W^{(2)}$ are standard
Brownian motions with $\langle W^{(1)}, W^{(2)}\rangle_t = \rho t$,
with $|\rho|<1$ constant. The process $(Y_t)$ is a fast mean-reverting
process with rate of mean reversion $1/\delta$ ($\delta>0$).
The parameters~$\beta$ and $\sigma(y)$ are chosen to satisfy the following.
%
%
\begin{assmpn}\label{assmpn1}
We assume that:
{\renewcommand\thelonglist{(\arabic{longlist})}
\renewcommand\labellonglist{\thelonglist}
\begin{longlist}
\item\label{Assmbeta}
$\beta\in\{ 0 \} \cup[ \frac12, 1)$;
\item\label{Ytpositive} in the case of $\beta=1/2$, we require
$m>\nu
^2/2$ and $Y_0>0$ a.s.,
in the case of $1/2<\beta<1$, we require $m>0$ and $Y_0>0$ a.s.;
\item\label{Assmsigma} $\sigma(y) \in C(\R; \R_+)$ satisfies
\[
\sigma(y) \leq C(1+|y|^\sigma)
\]
for some constants $C>0$ and $\sigma$ with $0 \leq\sigma<1-\beta$.
\end{longlist}
}
\end{assmpn}

These assumptions ensure existence and uniqueness of a strong solution
of~(\ref{sde-1}). This can be seen as
a combination of existence of martingale problem solution (e.g., Theorem
5.3.10 in Ethier and Kurtz \cite{EK85})
and the Yamada--Watanabe theory for 1-D diffusions (e.g., Chapter 5,
Karatzas and Shreve~\cite{KS99}).
In particular, Assumption \ref{assmpn1}\ref{Ytpositive} ensures that,
in the case $\beta\in[\frac12,1)$, $Y_t>0$ a.s. for all $t\geq0$
(see Appendix \ref{secFellerclassification}). In the case $\beta=0$,
$Y$ is an Ornstein--Uhlenbeck (OU) process with a natural state space
$(-\infty, \infty)$. In order to present both model cases using one
simple set of notation, we denote the state space for $Y$ as $E_0$ with
$E_0:=\R$ if $\beta=0$ and $E_0 :=(0,\infty)$ when $\beta\in[\frac12,1)$.

Note that the Heston model, for which $\beta=1/2$ and $\sigma
(y)=\sqrt
{y}$, does not satisfy these assumptions, but it has been treated
separately in \cite{FFF} by explicit computation of the moment
generating function.

The infinitesimal generator of the $Y$ process, when $\delta=1$, can be
identified with the following differential operator on the class of
smooth test functions vanishing off compact sets:
%
%
\begin{equation}\label{B-operator}
B := (m-y) \partial_y + \tfrac12\nu^2 |y|^{2\beta} \partial^2_{yy}.
\end{equation}
Following the general theory of 1-D diffusion (e.g., Karlin and
Taylor \cite{KT81}, page~221), we introduce the so called
scale and speed measure of the $(Y_t)$ process,
\[
s(y) := \exp\biggl\{-\int_1^y\frac{2(m-z)}{\nu^2|z|^{2\beta
}}\,dz \biggr\},\qquad
m(y) := \frac{1}{\nu^2|y|^{2\beta}s(y)}.
\]
Denoting $d S(y) :=s(y) \,dy$ and $dM(y) := m(y) \,dy$, we then have
%
%
\begin{equation}\label{decomp}
B f(y) = \frac12 \,\frac{d}{d M} \biggl[\frac{d f(y)}{d S} \biggr].
\end{equation}
Under Assumption \ref{assmpn1} there exists a unique probability measure
%
%
\begin{equation}\label{pi}
\pi(dy) := Z^{-1} m(y) \,dy, \qquad Z := \int_{E_0} m(y) \,dy <
\infty
\end{equation}
such that $\int B f \,d \pi=0$ for all $f \in C^2_c(E_0)$. See
Appendix \ref{Ypprocess}.

By a change of variable $X_t=\log S_t$, we have
\[
dX_t= \bigl(r-\tfrac{1}{2}\sigma^2(Y_t) \bigr)\,dt+\sigma(Y_t)\,dW^{(1)}_t.
\]
In order to study small time behavior of the system, we rescale time
$t\mapsto\ep t$ for $0<\ep\ll1$; denoting the rescaled processes by
$X_{\ep,\delta,t}$ and $Y_{\ep,\delta,t}$, we have, in distribution,
%
%
\begin{subequation}\label{sde}
%
%
\begin{eqnarray}
\label{X}
dX_{\ep,\delta,t}&=&\ep\biggl(r-\frac{1}{2}\sigma^2(Y_{\ep
,\delta
,t})
\biggr)\,dt+\sqrt{\ep} \sigma(Y_{\ep,\delta,t})\,dW^{(1)}_t,\\
\label{Y}
dY_{\ep,\delta,t}&=&\frac{\ep}{\d} (m-Y_{\ep,\delta,t}) \,dt +
\nu
\sqrt
{\frac{\ep}{\delta}} Y^\beta_{\ep,\delta,t}\,d W^{(2)}_t.
\end{eqnarray}
\end{subequation}

We are interested in understanding the two-scale $\varepsilon, \delta
\rightarrow0$ limit behavior of option prices and its implication to
implied volatility. In this paper, we restrict our attention to the
following two regimes:
\[
\d=\ep^4 \quad\mbox{and}\quad\d=\ep^2.
\]
In view of \cite{FFF}, to obtain a large deviation estimate of option
prices, it is sufficient to obtain a large deviation principle (LDP)
for $\{X_{\ep,\delta,t}\dvtx\ep>0\}$. By Bryc's inverse Varadhan lemma
\cite{DeZ98} (Theorem 4.4.2), we know that the key step is proving
convergence of the following functionals:
%
%
\begin{equation}\label{u-defn}
\ued(t,x,y):= \varepsilon\log E\bigl[ e^{\varepsilon^{-1} h(\Xedt)}
|X_{\varepsilon
,\delta,0}=x, Y_{\varepsilon,\delta,0}=y\bigr],\qquad h\in
C_b(\R),\hspace*{-40pt}
\end{equation}
to some quantity independent of $y$. The rate function in the LDP is
then given in terms of a variational formula involving the limit of the
functionals~$u_{\ep,\d}$.

For each $h\in C_b(\R)$, the function $\ued$ satisfies a nonlinear
partial differential equation given in (\ref{u-eqn}). In Section \ref
{heuristics}, we use heuristic arguments to obtain PDEs that
characterize the limit of these $\ued$. Proving this convergence
rigorously, however, is nontrivial. Intuitively we know that, as $Y$
has a mean reversion rate $1/\delta$ and $\delta\ll\ep$, the effect of
the $Y$ process should get averaged out. To be exact, the form of
nonlinear operator (\ref{Hep}) indicates that convergence of $\ued$
is an averaging problem (over the fast $y$ variable) for
Hamilton--Jacobi equations. Such problems, in the context of compact
state space for the averaging variable, can be handled
by extending standard linear equation techniques using viscosity
solution language. The $Y$ process in this article lies in $E_0$, which
is $\R$ in the case of $\beta=0$ and $(0,\infty)$ in other cases. $E_0$
is a noncompact space, and therein lies an additional difficulty.

We adapt methods developed in Feng and Kurtz \cite{FK06}. Indeed, an
abstract method for large deviation for sequence of Markov processes,
based on convergence of HJB equation, is developed fully in
\cite{FK06}. The two schemes treated in this article are of the nature
of Examples 1.8 and 1.9, introduced in Chapter 1, and proved in
detail in Chapter 11 of \cite{FK06}. In this article, we not only
present a direct proof, but also introduce some argument to further
simplify \cite{FK06} in the setting of multi-scale. This is possible in
a large part due to the~locally compact state space and mean-reverting
nature of the process~$Y$.\looseness=-1

In particular, modulo technical subtleties in verification of
conditions, the setup of Section 11.6 in \cite{FK06} corresponds to the
large deviation result in our case of \mbox{$\delta=\varepsilon^2$}.
Since $E_0$
is locally compact, and we only deal with PDEs instead of abstract
operator equations, great simplification of \cite{FK06} can be achieved
through the use of a special class of test functions. See
Conditions \ref{limsupH} and \ref{liminfH}.
The techniques we introduce (Lemmas \ref{sub-super-solns} and \ref
{rigconv}) are not limited to averaging problems, but are also
applicable to problems of homogenization, which we will not delve into
in this article. The rigorous justification of convergence of $\ued$ is
shown in Section \ref{rigorous}.

The main results of the paper are stated in Section \ref{mainresults}.
Theorem \ref{LDP} is a~rare event large deviation-type estimate
corresponding to short time, out-of-the-money option pricing. Corollary
\ref{option-price} and Theorem \ref{impliedvol} give asymptotics of
option price and implied volatility, respectively, for such situations.
The proofs are given in the sections that follow, starting with
heuristic proofs in Section \ref{heuristics} and finishing with
rigorous justifications in Sections \ref{resultsFK} and
\ref{rigorous}. The technical results in Lemmas \ref{sub-super-solns}
and \ref{rigconv} may be of independent interest.

\section{Main results}\label{mainresults}
Observe that in the SDE (\ref{sde}), while the scaled log stock price
process runs on a time scale of order $\ep$, the scaled $Y$ process
runs on a time scale of order $\ep/\delta$. This is due to the
extremely short mean-reversion time, $\delta=\ep^r$ ($r=2,4$), of the
$Y_{\ep,\delta,\cdot}$ process. Thus, as $\ep$ approaches zero,
long-time behavior of the unscaled $Y$ process comes into play. This
long-time behavior of the $Y$ process manifests itself in the large
deviation principle (LDP) of the scaled log stock price via the
quantities $\bar{\sigma}{}^2$ and $\bar{H}_0$ defined below. 
Define
%
%
\begin{equation}\label{sigmabar}
\bar{\sigma}^2 := \int\sigma^2(y) \pi(dy);
\end{equation}
the average of the volatility function $\sigma^2(\cdot)$ with respect
to the invariant distribution of $Y$.
Recall $B$, the generator of the $Y$ process, defined in (\ref
{B-operator}). Define the perturbed generator
%
%
\begin{equation}\label{pertB}
B^pg(y)=Bg(y)+\rho\sigma\nu y^\beta p \,\partial_y g(y), \qquad g\in
C^2_c(E_0).
\end{equation}
Let $Y^p$ be the process corresponding to generator $B^{p}$, and define
%
%
\begin{equation}\label{DV}
\bar{H}_0(p) := \limsup_{T \rightarrow+\infty} \sup_{y \in E_0}
T^{-1} \log E\bigl[ e^{(1/2) |p|^2 \int_0^T \sigma^2(Y^p_s) \,ds
}|Y^p_0=y\bigr].
\end{equation}
$Y^p$ has strong enough ergodic properties that the limit above does
not depend upon $y$ even if we omitted the $\sup_{y\in E_0}$; and, in
fact,\vspace*{1pt} the $\limsup_{T\to\infty}$ can be replaced with
$\lim_{T\to
\infty
}$ in the above definition. We will justify this fact in the rigorous
derivations.
By Girsanov's transformation
%
%
\begin{equation}\label{H0-alt}
\bar{H}_0(p) = \limsup_{T \rightarrow+\infty} T^{-1} \log E[
e^{\int
_0^T\rho p \sigma(Y_s) \,d W^{(2)}(s) + ({(1-\rho^2)}/{2})|p|^2 \int
_0^T \sigma^2(Y_s) \,ds} ] ,\hspace*{-35pt}
\end{equation}
where $Y$ is the process with generator $B$. From this expression, we
see that~$\bar{H}_0$ is convex and superlinear in $p$. $\bar{H}_0(p)$
is the scaled limit of the log moment generating function of a function
of occupation measures of the process $Y^p$. As such, it has an
equivalent representation in terms of the rate function for the LDP of
occupation measures of $Y^p$. This equivalent representation of
$\bar{H}_0$ is given in (\ref{H0var}) in Section \ref{delta2-rig}.

Having defined these crucial terms, we proceed to the statement of our
results.\vspace*{-2pt}
%
%
\begin{theorem}[(Large deviation)]\label{LDP}
Assume $X_{\ep,\ep^r,0}=x_0$ and $Y_{\ep,\ep^r,0}=y_0$ where $r=2,4$
and suppose that Assumption \ref{assmpn1} holds. For $x\in\R$, let
%
%
\begin{equation}
\label{I4}
I_4(x;x_0,t):=\frac{|x_0-x|^2}{2\bar{\sigma}^2t},
\end{equation}
where $\bar{\sigma}$ is defined in (\ref{sigmabar}) and
\begin{equation}
\label{I2}
I_2(x;x_0,t):=t\bar{L}_0 \biggl(\frac{x_0-x}{t} \biggr),
\end{equation}
where $\bar{L}_0$ is the Legendre transform of $\bar{H}_0$ defined in
(\ref{DV}). 

Then, for each regime $r\in\{2,4\}$, for every fixed $t>0$ and $x_0
\in\R, y_0 \in E_0$, a~large deviation principle (LDP) holds for $\{
X_{\ep,\ep^r,t}\dvtx\ep>0\}$ with speed $1/\ep$ and good rate function
$I_r(x; x_0,t)$. In particular,
%
%
\begin{equation}\label{LDP1}
\lim_{\ep\to0} \ep\log P(X_{\ep,\ep^r,t}>x)=-I(x;x_0,t)
\qquad\mbox
{when }x>x_0.
\end{equation}
Similarly, when $x<x_0$, we have
%
%
\begin{equation}\label{LDP2}
\lim_{\ep\to0} \ep\log P(X_{\ep,\ep^r,t}<x)=-I(x;x_0,t).\vspace*{-2pt}
\end{equation}
\end{theorem}
%
%
\begin{rem}
The rate functions $I_r(x;x_0,t)$, in both regimes, are convex,
continuous functions of $x$ and $I_r(x_0;x_0,t)=0$.\vspace*{-2pt}
\end{rem}
%
%
\begin{rem}
In the case $\delta=\ep^4$, observe that the rate function $I_4$,
in~(\ref{I4}), is the same as the rate function for the Black--Scholes
model with constant volatility~$\bar{\sigma}$.
In other words, in the ultra fast regime, to the leading order, it is
the same as averaging first and then taking the short maturity limit.\vspace*{-2pt}
\end{rem}
%
%
\begin{rem}
In the case $\delta=\ep^2$, no explicit formula for the rate function
is obtained. However, an explicit formula of the rate function is
obtained for the Heston model in \cite{FFF} which corroborates the
formula in (\ref{I2}). The Heston model per se does not fall in the
category of stochastic volatility models covered in this paper, but
direct computation of $\bar{H}_0$, given by (\ref{DV}) and $\bar{L}_0$,
its Legendre transform, is possible for this model.\vadjust{\goodbreak}
\end{rem}

Let $S_0>0$ be the initial value of stock price, and let $X_{\ep,\ep
^r,0}=x_0=\log S_0$. The asymptotic behavior of the price of
out-of-the-money European call option with strike price $K$ and short
maturity time $T=\ep t$ is given in the following corollary. We only
consider out-of-the-money call options by taking
%
%
\begin{equation}\label{OTM-cond}
S_0<K \quad\mbox{or}\quad x_0<\log K.
\end{equation}
The other case, $S_0>K$, is easily deduced by considering
out-of-the-money European put options and using put-call parity.
%
%
\begin{cor}[(Option price)]\label{option-price}
For fixed $t>0$,
\[
\lim_{\ep\to0^+}\ep\log E [e^{-r\ep t} (S_{\ep,\ep
^r,t}-K )^+
]=-I_r(\log K;x_0,t)
\]
for $r=2,4$.
\end{cor}

Denote the Black--Scholes implied volatility for out-of the-money
European call option, with strike price $K$, by $\sigma_{r,\ep
}(t,\log
K, x_0)$, where $r=2,4$ correspond to the two regimes.
By the same argument used in \cite{FFF}, we get an asymptotic formula
for implied volatility:
%
%
\begin{theorem}[(Implied volatilities)]\label{impliedvol}
\[
\lim_{\ep\to0^+}\sigma^2_{r,\ep}(t,\log K, x_0)=\frac{(\log
K-x_0)^2}{2I_r(\log K;x_0,t) t}.
\]
\end{theorem}
%
%
\begin{rem}
In the case $\delta=\ep^4$, the implied volatility is $\bar{\sigma}$,
which is obtained by averaging the volatility term $\sigma^2(y)$ with
respect to the equilibrium measure for~$Y$. It is likely that more
features of the $Y$ process, beyond its equilibrium, will be manifested
in higher order terms of implied volatility. Studying the next order
term of implied volatility is a topic for future research.
\end{rem}
%
%
\begin{rem}
The limit of at-the-money implied volatility, that is,\break $\lim_{\ep\to
0}\sigma^2_{r,\ep}(t,x_0, x_0)$, is obtained as in \cite{FFF}, Lemma
2.6. However, the continuity of the limiting implied volatility at
$\log K=x_0$ is not obvious in the $r=2$ case. We discuss this at the
end of Section \ref{impliedvolatility}.
\end{rem}

\section{Preliminaries}\label{prelim}
The process $(X_{\varepsilon,\delta}, Y_{\varepsilon,\delta})$ is Markovian,
and can be identified through a martingale problem given by generator
%
%
\begin{eqnarray}
\Aed f(x,y) &=& \varepsilon\biggl( \biggl( r - \frac12 \sigma^2(y)\biggr) \,\partial_x
f(x,y) + \frac12 \sigma^2(y) \,\partial^2_{xx} f(x,y)\biggr)
\nonumber\\[-8pt]\\[-8pt]
&&{} + \frac{\varepsilon}{\delta}B f(x,y) + \frac{\varepsilon
}{\sqrt
{\delta}} \rho\sigma(y) \nu y^\beta\,\partial^2_{xy} f(x,y),
\nonumber
\end{eqnarray}
where $f \in C^2_c(\R\times E_0)$. Recall that $B$ is given by (\ref
{B-operator}).
Let $g \in C_b(\R)$ and define
%
%
\begin{equation}\label{Kac-rep}
v_{\ep,\d}(t,x,y):= E [ g(\Xedt)|X_{\ep,\d,0}=x,Y_{\ep,\d
,0}=y ].
\end{equation}
In general, $v_{\ep,\d}\in C_b([0,T]\times\R\times E_0)$. If,
moreover, $v_{\ep,\d}\in C^{1,2}([0,T]\times\R\times\R)$, then it
solves the following Cauchy problem in classical sense:
%
%
\begin{subequation}\label{v-eqn}
%
%
\begin{eqnarray}
\partial_tv&=&\Aed v\qquad\mbox{in } (0,T]\times\R\times E_0;\\
v(0,x,y)&=&g(x),\qquad (x,y)\in\R\times E_0.
\end{eqnarray}
\end{subequation}

\subsection{Logarithmic transformation method}
Recall the definition of $u_{\ep,\d}$ in~(\ref{u-defn}). That is,
$u_{\varepsilon,\delta} :=\varepsilon\log v_{\varepsilon,\delta}$ when
$g(x)=e^{\ep^{-1}h(x)}, h\in C_b(\R)$, in~(\ref{Kac-rep}). By~(\ref
{v-eqn}) and some calculus, at least informally,~(\ref{u-eqn}) below is
satisfied. This is the logarithmic transform method by Fleming and
Sheu. See Chapters~VI and VII in \cite{FS06}. In general, in the absence of
knowledge on smoothness of~$v_{\varepsilon,\delta}$, we can only conclude
that $u_{\ep,\d}$ solves the Cauchy problem (\ref{u-eqn}) in the sense
of viscosity solution (Definition~\ref{vdef}). In addition to Fleming
and Soner~\cite{FS06}, such arguments can also be found in
Section 5 of Feng~\cite{Fe99}.
%
%
\begin{lemma}
For $h\in C_b(\R)$, $\ued$ defined as in (\ref{u-defn}), is a bounded
continuous function satisfying the following nonlinear Cauchy problem
in viscosity solution sense:
%
%
\begin{subequation}\label{u-eqn}
%
%
\begin{eqnarray}
\partial_t u&=&\Hed u\qquad \mbox{in }(0,T]\times\R\times E_0; \\
u(0,x,y)&=&h(x),\qquad (x,y)\in\R\times E_0.
\end{eqnarray}
\end{subequation}
In the above,
%
%
\begin{eqnarray}\label{Hep}
H_{\varepsilon,\delta}u(t,x,y)& = &\varepsilon e^{-\varepsilon^{-1}u}
\Aed
e^{\varepsilon^{-1}u}(t,x,y) \nonumber\\
&=& \varepsilon\biggl( \biggl( r - \frac12 \sigma^2(y)\biggr) \,\partial_x u+ \frac12
\sigma^2(y) \,\partial^2_{xx} u\biggr)\nonumber\\[-8pt]\\[-8pt]
&&{}+ \frac12 | \sigma(y) \,\partial
_x u|^2
+ \frac{\varepsilon^2}{\delta} e^{- \varepsilon^{-1} u} B
e^{\varepsilon
^{-1} u}\nonumber\\
&&{} + \rho\sigma(y) \nu y^\beta\biggl( \frac{\varepsilon}{\sqrt
{\delta}}\,
\partial^2_{xy} u + \frac{1}{\sqrt{\delta}} \,\partial_x u \,\partial
_y u\biggr),
\nonumber
\end{eqnarray}
where
\[
\frac{\varepsilon^2}{ \delta} e^{-\varepsilon^{-1} u} B
e^{\varepsilon^{-1} u}
= \frac{\varepsilon}{\delta} B u + \delta^{-1} \frac12 |\nu
y^\beta\,
\partial_y u|^2 .
\]
Note that $\Hed$ only operates on the spatial variables $x$ and $y$.
\end{lemma}

\subsection{Heuristic expansion} \label{heuristics}
By Bryc's inverse Varadhan lemma (e.g., Theorem~4.4.2 of \cite{DeZ98}),
we know that convergence of $\ued$ is a necessary condition to obtain
the LDP for $\{\Xedt\dvtx\ep>0\}$. In this section, we describe
heuristical\-ly PDEs characterizing $\ued$ in the limit and the nature of
convergence itself.\vadjust{\goodbreak}

Henceforth, for notational simplicity, we will drop the subscript
$\delta$ and write $\u$ and $\H$ for $\ued$ and $\Hed$,
respectively. We
begin by the following heuristic expansion of $\u$ in integer powers of
$\ep$:
%
%
\begin{equation}\label{heur-u}
\u=u_0+\ep u_1+\ep^2 u_2+\ep^3 u_3+\ep^4u_4+\cdots
\end{equation}
in both regimes. The $u_i, i=0,1,\ldots,$ are functions of $t,x,y$. In
this heuristic section, we make reasonable choices of $u_i$ which a
posteriori, following a~rigorous proof of the convergence of $u_\ep$ in
Section \ref{rigorous}, are shown to be the right choice.

\subsubsection{\texorpdfstring{The case of $\delta= \varepsilon^4$}{The case of delta = epsilon 4}}\label{heurep4}

Computation of $H_\varepsilon u_\varepsilon$ [see (\ref{Hep})]
reveals that,
in this scale, the fast process $Y$ oscillates so fast that averaging
occurs up to terms of order $\varepsilon^2$. Namely, $u_0=u_0(t,x)$,
$u_1=u_1(t,x)$ and $u_2=u_2(t,x)$ will not depend on $y$. To see this,
we equate coefficients of powers of $\ep$ in $\partial_t\u=\H\u$.

Terms of $O (\frac{1}{\ep^4} )$ satisfy
\[
0=\tfrac1 2 \nu^2y^{2\beta}(\partial_y u_0)^2,
\]
so we choose $u_0$ independent of $y$. With this choice of $u_0$ the
equation for the coefficients of the next order terms, which is of
$O
(\frac{1}{\ep^2} )$, reduces to
\[
0=Bu_1+\tfrac1 2 \nu^2y^{2\beta}(\partial_y u_1)^2.
\]
This equation is satisfied by choosing $u_1$ independent of $y$. With
this choice of $u_1$, the equation for coefficients of the next order
terms, of $O (\frac{1}{\ep} )$, becomes
\[
0=Bu_2.
\]
By choosing $u_2$ independent of $y$ the last equation is satisfied.

Thus, by these choices of $u_0, u_1$ and $u_2$ independent of $y$, it
follows that
\begin{eqnarray*}
H_\varepsilon u_\varepsilon(x,y) &=& \tfrac12 |\sigma(y) \,\partial_x u_0|^2
+ B
u_3 \\
&&{} + \varepsilon\bigl( \sigma^2(y) \,\partial_x u_0 \,\partial_x u_1 + \tfrac12
\sigma^2(y) \,\partial_{xx} u_0\\
&&\hspace*{20pt}{}  + \bigl(r-\tfrac12 \sigma^2(y)\bigr) \,\partial_x u_0
+ \nu\rho\sigma(y) y^\beta\,\partial_x u_0 \,\partial_y
u_3 +
B u_4 \bigr) \\
&&{} + o(\varepsilon).
\end{eqnarray*}

The $\varepsilon^0$ order terms then satisfy
%
\[
\partial_t u_0(t,x)=\tfrac12 | \partial_x u_0(t, x)|^2 \sigma^2(y) + B
u_3(t, x,y), 
\]
that is,
\[
B u_3(t, x,y) = \partial_t u_0(t,x) -\tfrac12 | \partial_x u_0(t, x)|^2
\sigma^2(y).
\]
The above is a Poisson equation for $u_3$ with respect to the operator
$B$ in the $y$ variable. We impose the condition that the right-hand
side is centered with respect to the invariant distribution $\pi$
[given in (\ref{pi})]. This ensures a~solution to the Poisson equation,
which is unique up to a constant in\vadjust{\goodbreak} $y$. See Appendix
\ref{growthestimates} for growth estimates of the solution.
Therefore we get
\[
\partial_t u_0(t,x) = \tfrac12 |\bar{\sigma}\,\partial_x u_0(t,x)|^2;
\]
where
\[
\bar{\sigma}^2 = \int\sigma^2(y) \pi(dy).
\]
Thus the leading order term in the heuristic expansion satisfies
%
%
\begin{subequation}\label{u0-H1}
%
%
\begin{eqnarray}
\partial_t u_0&=& \bar{H}_0 u_0(x) ,\qquad t>0 ; \\ 
u_0(0,x)&=&h(x),
\end{eqnarray}
\end{subequation}
where
\[
\bar{H}_0 u_0(x) := \tfrac12 |\bar{\sigma}\,\partial_x u_0(x)|^2.
\]

\subsubsection{\texorpdfstring{The case of $\delta=\varepsilon^2$}{The case of delta = epsilon 2}}\label{heurep2}

When $\delta$ goes to zero at a slower rate $\varepsilon^2$, limits
become very different and more features in the $Y$ process (rather than
just its equilibrium) is retained. We observe that while $u_0$ is
independent of $y$ as in the faster scaling regime, $u_1$ may now
depend on $y$. Equating coefficients of $O (\ep^{-2} )$
in $\partial
_t\u=\H\u$ we get
\[
0=\tfrac1 2 \nu^2y^{2\beta}(\partial_y u_0)^2,
\]
and so we choose $u_0=u_0(t,x)$ independent of $y$. Then $H_\ep u_\ep$
reduces to
\begin{eqnarray*}
H_\varepsilon u_\varepsilon(t,x,y) &=& \tfrac12 |\sigma(y) \,\partial_x
u_0|^2 +
\rho\sigma(y)\nu y^\beta\,\partial_x u_0 \,\partial_y u_1 + e^{-u_1} B
e^{u_1} \\
& &{} + \varepsilon\bigl( \sigma^2(y) \,\partial_x u_0 \,\partial_x u_1 +
\tfrac
12 \sigma^2(y) \,\partial_{xx} u_0 + \bigl(r-\tfrac12 \sigma^2(y)\bigr) \,\partial_x
u_0 \\
& &\hspace*{20.6pt}{} +B u_2 + \nu y^{2\beta}\,\partial_yu_1\,\partial_yu_2+ \rho
\sigma(y)\nu y^\beta\,\partial_{xy} u_1\\
& &\hspace*{55pt}{}  +\rho\sigma(y)\nu y^\beta
\,\partial_x u_1 \,\partial_y u_1 + \rho\sigma(y)\nu y^\beta\,\partial_x u_0 \,\partial_y
u_2\bigr) \\
& &{} + o(\varepsilon).
\end{eqnarray*}
The leading order terms should satisfy
%
%
\begin{eqnarray}\label{H0-version1}\quad
\partial_t u_0(t,x)&=&\tfrac12 |\partial_x u_0(t,x)|^2 \sigma^2(y) +
\rho
\nu\sigma(y) y^\beta\,\partial_x u_0(t,x) \,\partial_y u_1(t,x,y)
\nonumber\\[-8pt]\\[-8pt]
&&{} + e^{-u_1}B e^{u_1}(t,x,y). \nonumber
\end{eqnarray}
%
We will rewrite the above equation as an eigenvalue problem.
Recall $B$, the generator of the $Y$ process defined in (\ref
{B-operator}) and the perturbed generator $B^p$ defined in~(\ref{pertB}).
Then
%
%
\begin{equation}\label{pertB-psi}
e^{-u_1} B e^{u_1} + \rho\sigma(y)\nu y^\beta\,\partial_x u_0
\,\partial
_y u_1
= e^{-u_1} B^{\partial_x u_0(t,x)} e^{u_1}.
\end{equation}

Fix $t$ and $x$, and rewrite (\ref{H0-version1}) %
in terms of the perturbed generator (\ref{pertB-psi}).
\[
e^{-u_1} B^{\partial_x u_0(t,x)} e^{u_1}(t,x,y)+\tfrac12 |\partial_x
u_0(t,x)|^2 \sigma^2(y) =\partial_tu_0(t,x).
\]
Multiplying the above equation by $e^{u_1}$, we get the eigenvalue problem
%
%
\begin{equation}\label{EV-problem}
(B^{\partial_x u_0}+V )g(y)=\lambda g(y),
\end{equation}
where $V(\cdot)=\frac12 |\partial_x u_0(t,x)|^2 \sigma^2(\cdot) $
is a
multiplicative potential operator, $g(\cdot)=e^{u_1(t,x,\cdot)}$ and
$\lambda(t,x)=\partial_t u_0(t,x)$.
Choose $u_1$ such that $(\lambda, g )$ is the solution to the principal
(positive) eigenvalue problem (\ref{EV-problem}).
Note that the dependence of the eigenvalue, $\lambda$, on $t$ and $x$
is only through $\partial_x u_0$. If (\ref{EV-problem}) can be solved
with a nice~$g$, then 
we have
%
%
\begin{equation}\label{H0bar}
\lambda(t,x)= \bar{H}_0(\partial_x u_0), 
\end{equation}
where $\bar{H}_0$ is defined as (\ref{DV}). The leading order terms
then satisfy
%
%
\begin{equation}\label{u000-1}
\partial_t u_0(t,x) = \bar{H}_0 (\partial_x u_0(t,x)).
\end{equation}

Constructing a classical solution for (\ref{EV-problem}) is a
considerably hard problem, even in the 1-D situation. If (\ref
{EV-problem}) can be solved with a nice $g$, then (\ref{DV}) always
holds with the $\bar{H}_0$ given by (\ref{H0bar}). The converse is not
always true. Especially, (\ref{DV}) says nothing about the
eigenfunction $g$. However, we only need the definition in (\ref{DV})
in rigorous treatment of the problem. We will show (in Section
\ref{delta2-rig}) that (\ref{u000-1}) is the limit equation where
$\bar
{H}_0$ is given by (\ref{DV}) irrespective of whether a solution to the
eigenvalue problem (\ref{EV-problem}) exists or does not.

To summarize,
%
%
\begin{subequation}\label{u000}
%
%
\begin{eqnarray}
\partial_t u_0(t,x) &=& \bar{H}_0 (\partial_x u_0(t,x)),\qquad t>0;\\
u_0(0,x)&=&h(x),
\end{eqnarray}
\end{subequation}
where $\bar{H}_0$ is given by (\ref{DV}) or (\ref{H0-alt}).

\section{Convergence of HJB equations} \label{resultsFK}
The results of this section can be independently read from the rest of
the article.

We reformulate and simplify some techniques, regarding multi-scale
convergence of HJB equations, introduced in \cite{FK06}.
Compared with \cite{FK06}, the simplification makes ideas more
transparent and readily applicable. These are made possible because
we are dealing with Euclidean state spaces which are locally compact.
All these results are generalizations of Barles--Perthame's
half-relaxed limit argument first introduced in single scale, compact
state space
setting.

Let $E\subset\R^m$, $E_0 \subset\R^n$ and $E^\prime:=E \times E_0
\subset\R^d$ where $d=m+n$. A typical element in $E$ is denoted as
$x$, and
a typical element in $E^\prime$ is denoted as $z=(x,y)$ with $x \in E$
and $y \in E_0$.
We denote a class of compact sets in $ E^\prime$
\[
\mathcal Q := \{ K \times\tilde{K} \mbox{: compact } K \subset
\subset
E \mbox{, compact } \tilde{K} \subset\subset E_0 \}.
\]
We specify a family of differential operators next.
Let $\Lambda$ be an index set and
\begin{eqnarray*}
H_i(x, p, P; \alpha)\dvtx E \times\R^m \times M_{m\times m} \times
\Lambda&\mapsto&\R,\qquad i=0,1; \\
H_\varepsilon(z, p,P) \dvtx E^\prime\times\R^d \times M_{d \times d}
&\mapsto&\R
\end{eqnarray*}
be continuous. For each $f \in C^2(\R^d)$, let $\nabla f(x) \in\R^d$
and $D^2f(x) \in M_{d \times d}$, respectively, denote gradient
and Hessian matrix evaluated at $x$.
We consider a sequence of differential operators
\[
H_\varepsilon f(z) : = H_\varepsilon(z, \nabla f(z), D^2f(z))
\]
for $f$ belongs to the following two domains:
\begin{eqnarray*}
D_{\varepsilon, +} &:=& \{ f \dvtx f \in C^2(E^\prime), f \mbox{ has compact
finite level sets} \}; \\
D_{\varepsilon, -} &:=& -D_{\varepsilon,+}:= \{ - f \dvtx f \in
C^2(E^\prime), f
\mbox{ has compact finite level sets} \}.
\end{eqnarray*}
We will separately consider these two domains depending on the
situation of sub- or super-solution.
We also define domains $D_+, D_-$ similarly replacing~$E^\prime$ by $E$.

We will give conditions where $u_\varepsilon(t,z)=u_\varepsilon
(t,x,y)$ solving
%
%
\begin{equation}\label{sol}
\partial_tu_\varepsilon(t,z)=H_\varepsilon(z, \nabla u_\varepsilon
(t,z), D^2
u_\varepsilon(t,z))
\end{equation}
converging to $u(t,x)$ which is a sub-solution to
%
%
\begin{equation}\label{sub}
\partial_tu(t,x)\leq\inf_{\alpha\in\Lambda} H_0(x,\nabla u(t,x), D^2
u(t,x);\alpha)
\end{equation}
and a super-solution to
%
%
\begin{equation}\label{super}
\partial_tu(t,x)\geq\sup_{\alpha\in\Lambda} H_1(x,\nabla u(t,x),
D^2 u(t,x);\alpha).
\end{equation}
The meaning of sub- super-solutions is defined as follows (as, e.g.,
in Fleming and Soner \cite{FS06}).
%
%
\begin{definition}[(Viscosity sub- super-solutions)]\label{vdef}
We call a bounded measurable function $u$ a viscosity sub-solution to
(\ref{sub}) [resp., super-solution to (\ref{super})],
if $u$ is upper semicontinuous (resp., lower semicontinuous), and for each
\[
u_0(t,x)=\phi(t)+f_0(x),\qquad \phi\in C^1(\R_+), f_0 \in D_+,
\]
and each $x_0 \in E$ satisfying $u - u_0$ has a local maximum [resp., each
\[
u_1(t,x)=\phi(t)+f_1(x),\qquad \phi\in C^1(\R_+),f_1 \in D_-,
\]
and each $x_0 \in E$ satisfying $u - u_1$ has a local minimum] at
$x_0$, we have
\[
\partial_tu_0(t_0,x_0)-\inf_{\alpha\in\Lambda} H_0(x_0, \nabla
u_0(t_0,x_0), D^2 u_0(t_0,x_0); \alpha) \leq0,
\]
respectively,
\[
\partial_tu_1(t_0,x_0)-\sup_{\alpha\in\Lambda} H_1(x_0, \nabla u_1
(t_0, x_0), D^2 u_1(t_0,x_0); \alpha) \geq0.
\]

If a function is both a sub- as well as a super-solution, then it is a
solution.\vadjust{\goodbreak}
\end{definition}

We will assume the following two conditions.
%
%
\begin{condition}[(limsup convergence of operators)]\label{limsupH}
For each $f_0 \in D_+$ and each $\alpha\in\Lambda$, there exists
$f_{0,\varepsilon} \in D_{\varepsilon,+}$ (may depend on $\alpha$)
such that:
{\renewcommand\thelonglist{(\arabic{longlist})}
\renewcommand\labellonglist{\thelonglist}
\begin{longlist}
\item\label{Kf0}
for each $c>0$, there exists $K \times\tilde{K}
\in\mathcal Q$ satisfying
\[
\{ (x,y) \dvtx H_\varepsilon f_{0,\varepsilon} (x,y) \geq-c \} \cap
\{ (x,y)
\dvtx
f_{0,\varepsilon}(x,y) \leq c \}\subset K \times\tilde{K};
\]
\item for each $K \times\tilde{K} \in\mathcal Q$,
%
%
\begin{equation}\label{cond-b}{ \lim_{\varepsilon\rightarrow0} \sup
_{(x,y) \in K \times\tilde{K}}} |f_{0,\varepsilon} (x,y) - f_0(x) | =0;
\end{equation}
\item whenever $(x_\varepsilon, y_\varepsilon) \in K \times
\tilde{K}
\in\mathcal Q$ satisfies $x_\varepsilon\rightarrow x$,
%
%
\begin{equation}\label{cond-c}
\limsup_{\varepsilon\rightarrow0} H_\varepsilon f_{0,\varepsilon
}(x_\varepsilon,
y_\varepsilon) \leq H_{0}(x, \nabla f_0 (x), D^2 f_0(x); \alpha).
\end{equation}
\end{longlist}}
\end{condition}
%
%
\begin{condition}[(liminf convergence of operators)]\label{liminfH}
For each $f_1 \in D_{-}$ and each $\alpha\in\Lambda$, there exists
$f_{1,\varepsilon} \in D_{\varepsilon,-}$ (may depend on $\alpha$)
such that:
{\renewcommand\thelonglist{(\arabic{longlist})}
\renewcommand\labellonglist{\thelonglist}
\begin{longlist}
\item\label{Kpact} for each $c>0$, there exists $K \times
\tilde
{K} \in\mathcal Q$ satisfying
\[
\{ (x,y) \dvtx H_\varepsilon f_{1,\varepsilon} (x,y) \leq c \}
\cap\{ (x,y) \dvtx f_{1,\varepsilon}(x,y) \geq- c \}\subset K \times
\tilde
{K} ;
\]
\item for each $K \times\tilde{K}\in\mathcal Q$,
\[
{\lim_{\varepsilon\rightarrow0} \sup_{(x,y) \in K \times\tilde{K}}}
|f_1(x) - f_{1,\varepsilon} (x,y) | =0;
\]
\item whenever $(x_\varepsilon, y_\varepsilon) \in K \times
\tilde{K}
\in\mathcal Q$, and $x_\varepsilon\rightarrow x$,
\[
\liminf_{\varepsilon\rightarrow0} H_\varepsilon f_{1,\varepsilon
}(x_\varepsilon,
y_\varepsilon) \geq H_{1}(x, \nabla f_1 (x), D^2 f_1(x); \alpha).
\]
\end{longlist}}
\end{condition}

Let $u_\varepsilon$ be the viscosity solutions to (\ref{sol}); we define
\begin{eqnarray*}
u_3(t,x)&: =&
\sup\Bigl\{ \limsup_{\varepsilon\rightarrow0+} u_\varepsilon(t_\ep,
x_\varepsilon,
y_\varepsilon) \dvtx \exists(t_\ep, x_\varepsilon, y_\varepsilon)
\in
[0,T]\times K \times\tilde{K}, \\
&&\hspace*{126pt} (t_\ep, x_\varepsilon) \rightarrow
(t,x), K \times\tilde{K} \in\mathcal Q \Bigr\},
\\
{u}_4(t,x)&:=&
\inf\Bigl\{ \liminf_{\varepsilon\rightarrow0+} u_\varepsilon(t_\ep,
x_\varepsilon
, y_\varepsilon) \dvtx\exists(t_\ep, x_\varepsilon, y_\varepsilon)
\in
[0,T]\times K \times\tilde{K}, \\
&&\hspace*{119.5pt}(t_\ep, x_\varepsilon)
\rightarrow
(t,x), K \times\tilde{K} \in\mathcal Q \Bigr\},
\end{eqnarray*}
and $\bar{u} = u_3^*$ the upper semicontinuous regularization of
$u_3$ and $\underline{u} = (u_4)_*$ the lower semicontinuous
regularization of $u_4$.
%
%
\begin{lemma}\label{sub-super-solns}Suppose that $ \sup_{\varepsilon>0}
\Vert u_\varepsilon\Vert_\infty< \infty$.
Then:
\begin{longlist}[(2)]
\item[(1)] under Condition \ref{limsupH}, $\bar{u}$ is a
sub-solution to (\ref{sub});
\item[(2)] under Condition \ref{liminfH}, $\underline{u}$ is a
super-solution to (\ref{super}).\vadjust{\goodbreak}
\end{longlist}
\end{lemma}
\begin{pf}
Let $u_0(t,x) = \phi(t)+ f_0(x)$ for a fixed $\phi\in C^1(\R_+)$ and
$f_0 \in D_+$. Let $(t_0,x_0)$ be a local maximum of $\bar{u} -
u_0$, $t_0>0$.
We can modify $f_0$ and~$\phi$ if necessary so that $(t_0,x_0)$ is a
strict global maximum, for instance, by taking $\tilde{f}_0(x) = f_0(x)
+ k|x-x_0|^4$ and $\tilde{\phi}(t)= \phi(t) + k |t-t_0|^2$ for $k>0$
large enough.
Note that such modification has the property that
\[
{\lim_{\varepsilon\rightarrow0+}} \sup_{|x - x_0|<\varepsilon} |
\nabla
\tilde{f}_0(x) - \nabla f_0(x_0)| +
| D^2 \tilde{f}_0(x) - D^2 f_0(x_0)| = 0.
\]
Let $\tilde{u}_0 = \tilde{\phi} + \tilde{f}_0$.

Let $\alpha\in\Lambda$ be given. We now take $u_{0,\varepsilon} (t,z)=
\tilde{\phi}(t) + f_{0,\varepsilon}(z)$ where $f_{0,\varepsilon}$
is the
approximate of $\tilde{f}_0$ in Condition \ref{limsupH}.
Since $u_{\varepsilon}$ is bounded, and $u_{0,\varepsilon}$
has compact level sets, there exists $(t_\varepsilon, z_\varepsilon)
\in
[0,T] \times E^\prime$ such that
%
%
\begin{equation}\label{uemax}
( {u}_\varepsilon- u_{0,\varepsilon}) (t_\varepsilon, z_\varepsilon
) \geq(
{u}_\varepsilon- u_{0,\varepsilon}) (t, z) \qquad\mbox{for }(t,z)\in
[0,T]\times E^\prime
\end{equation}
and
%
%
\begin{equation}\label{Hsubineq}
\partial_t \tilde{\phi}(t_\varepsilon) - H_\varepsilon
f_{0,\varepsilon
}(z_\varepsilon) \leq0.
\end{equation}
The above implies $\inf_\varepsilon H_\varepsilon f_{0,\varepsilon
}(z_\varepsilon)
> -\infty$. We verify next that
$f_{0,\varepsilon}(z_\varepsilon) < c <\infty$. Then by Condition
\ref
{limsupH}\ref{Kf0}, there exists $K \times\tilde{K} \in\mathcal Q$
such that $z_\varepsilon=(x_\varepsilon, y_\varepsilon) \in K \times
\tilde{K}$.

Take a $(\hat{t}, \hat{x})$ such that $\tilde{u}_0(\hat{t}, \hat{x})
<\infty$. Take $\hat{z} = (\hat{x}, \hat{y})$ for some $\hat{y}
\in
E_0$. Then
\[
u_{0,\varepsilon}(\hat{t}, \hat{z}) = \tilde{\phi}(\hat{t}) +
f_{0,\varepsilon
}(\hat{z})
\rightarrow\tilde{\phi}(\hat{t}) +f_0(\hat{x}) = \tilde{u}_0(\hat{t},
\hat{x}) <\infty.
\]
Combined with (\ref{uemax}),
\[
u_{0,\varepsilon}(t_\varepsilon, z_\varepsilon) \leq2 \sup
_{\varepsilon>0}
\Vert
u_\varepsilon\Vert_\infty+ \sup_{\varepsilon>0} u_{0,\varepsilon
}(\hat{t},
\hat{z}) < \infty,
\]
and $\sup_{\varepsilon>0} f_{0,\varepsilon}(z_\varepsilon) <\infty
$ follows.

Since $K \times\tilde{K}$ is compact in $E^\prime$, there exists a
subsequence of $\{(t_\ep,z_\ep)\}$ (to simplify, we still use the
$\varepsilon$ to index it) and a $(\tilde{t}_0, \tilde{x}_0) \in[0,T]
\times E$ such that $t_\varepsilon\rightarrow\tilde{t}_0$
and $x_\varepsilon\rightarrow\tilde{x}_0$.
Such $(\tilde{t}_0, \tilde{x}_0)$ has to be the unique global maximizer
$(t_0, x_0)$ for $\bar u - \tilde{u}_0$ that appeared earlier.
This is because, by using
$x_\varepsilon\rightarrow\tilde{x}_0$ and $z_\varepsilon=
(x_\varepsilon,
y_\varepsilon)$, the definition of $\bar u$ and (\ref{cond-b}),
from
(\ref{uemax}) we have
%
%
\begin{equation}
(\bar{u} - u_0) (\tilde{t}_0, \tilde{x}_0) \geq(\bar{u} -
u_{0 }) (t, x)\qquad \forall(t,x).
\end{equation}

Now, from (\ref{Hsubineq}) and (\ref{cond-c}), we also have
\[
\partial_t u_{0}(t_0, x_0) \leq H_{0}(x_0, \nabla f_0 (x_0), D^2
f_0(x_0); \alpha).
\]
Note that $t_0, x_0$ and $u_0$ are all chosen prior to, and independent
of, $\alpha$. We can
take $\inf_{\alpha\in\Lambda}$ on both sides to get
\[
\partial_t u_{0}(t_0, x_0) - \inf_{\alpha\in\Lambda}H_{0}(x_0,
\nabla
u_0 (t_0,x_0), D^2 u_0(t_0, x_0); \alpha) \leq0.
\]

The proof that $\underline{u}$ is a super-solution of (\ref{super})
under Condition \ref{liminfH} follows similarly.
\end{pf}
%
%
\begin{lemma}\label{rigconv}
Suppose that the conditions in Lemma \ref{sub-super-solns} hold and
that there exists $h \in C_b(E)$ such that
\[
\lim_{\varepsilon\rightarrow0} \sup_{(x,y) \in K \times\tilde{K}} |
h(x) - u_\varepsilon(0, x,y) | =0\qquad \forall K \times\tilde{K} \in
\mathcal Q.
\]
Further suppose that for any sub-solution $u_0(t,x)$ of (\ref{sub})
with $u_0(0,x) = h(x)$ and
super-solution $u_1$ of (\ref{super}) with $u_1(0,x) = h(x)$, we have
\[
u_0(t,x) \leq u_1(t,x),\qquad (t,x) \in[0,T] \times E.
\]
That is, a comparison principle holds for sub-solutions of (\ref{sub})
and super-solutions of (\ref{super}) with initial data $h$.

Then $u= \bar{u} = \underline{u}$ and
\[
{\lim_{\ep\rightarrow0} \sup_{t \in[0,T]} \sup_{(x,y) \in K \times
\tilde{K}}} | u(t,x) - u_\varepsilon(t, x,y) | =0\qquad \forall K \times
\tilde{K} \in\mathcal Q.
\]
\end{lemma}

\section{Rigorous justification of expansions}\label{rigorous}
To rigorously prove the convergence of operators $\H$ given by (\ref
{Hep}) to operators $\bar{H}_0$ obtained by heuristic arguments in
Section \ref{heuristics}, we rely on and extend results developed in
\cite{FK06}. An exposition of the relevant results from \cite{FK06} was
laid out in Section \ref{resultsFK}.
In this section we verify Conditions \ref{limsupH} and \ref{liminfH}
and prove the comparison principle in Lemma \ref{rigconv}. We will
adhere to the notation used in Section \ref{resultsFK}.

Conditions \ref{limsupH} and \ref{liminfH} require us to carefully
choose a class of perturbed test functions with
an index set $\Lambda$ and a family of operators $ \{
H_0(\cdot;\alpha), H_1(\cdot;\alpha);\allowbreak \alpha\in\Lambda\}$ to obtain
viscosity sub- and super-solution estimates of $u_0$, the limit of
$u_\ep$.
This technique was first introduced in \cite{FK06}
and illustrated through examples in Chapter 11 of that book. Our
presentation simplifies the technique in the context of
application here.
We will make the sub-solution estimate given by $H_0(\cdot, \alpha)$
tight, by inf-ing over $\alpha$, hence introducing
yet another operator~$H_0$. Similarly, we sup over $\alpha$ to tighten
up the super-solution type estimate provided by
$H_1(\cdot,\alpha)$ which introduces operator $H_1$.

Let
%
%
\begin{equation}\label{zeta}
\zeta(y):= |y-m|^\zeta,
\end{equation}
where $\zeta>0$ is any number satisfying
$2\sigma<\zeta<2(1-\beta)$ with $\sigma$ and $\beta$ given as in
Assumption \ref{assmpn1}.
Throughout the two regimes ($\delta=\varepsilon^4, \varepsilon^2$),
we take
the index set
\[
\Lambda:= \{ \alpha= (\xi, \theta) \dvtx\xi\in C^2_c(E_0),
0<\theta<1
\};
\]
and define two domains
\[
D_+ := \{ f\dvtx f(x) = \varphi(x) + \gamma\log(1+ |x|^2) ; \varphi
\in
C^2_c(\R), \gamma>0\}
\]
and
\[
D_- := \{f: f(x) = \varphi(x) - \gamma\log(1+ |x|^2) ; \varphi\in
C^2_c(\R), \gamma>0 \}.
\]
A collection of compact sets in $\R\times E_0$ is defined by
\[
\mathcal Q:= \{ K \times\tilde{K} \mbox{: compact } K\subset
\subset\R
, \tilde{K} \subset\subset E_0 \}.\vadjust{\goodbreak}
\]

\subsection{\texorpdfstring{Case $\delta=\ep^4$}{Case delta = epsilon 4}}
For each $f=f(x) \in D_+$, and each $\alpha=(\xi, \theta) \in
\Lambda$,
we let
\[
g(y) := \xi(y) + \theta\zeta(y)
\]
and define perturbed test function
\[
f_\varepsilon(x,y) := f(x) + \varepsilon^3 g(y) = f(x) + \varepsilon
^3 \xi(y)
+ \varepsilon^3 \theta\zeta(y).
\]
Note that $\Vert\partial_x f \Vert_\infty+ \Vert\partial^2_{xx} f
\Vert_\infty< \infty$.
Then
\begin{eqnarray*}
\H f_\ep(x,y)&=&\ep\bigl[\bigl(r-\tfrac12\sigma^2(y)\bigr)\,\partial_x
f+\tfrac12\sigma
^2(y)\,\partial_{xx}^2f \bigr]+\tfrac12\sigma^2(y)|\partial_xf|^2\\
&&{} +B\xi(y)+\theta B\zeta(y)+\tfrac12\ep^2 \nu^2y^{2\beta
}|\partial
_y\xi(y)+\theta\,\partial_y\zeta(y)|^2\\
&&{} +\ep\rho\sigma(y)\nu y^\beta\,\partial_x f \bigl( \partial
_y\xi(y)+
\partial_y\zeta(y)\bigr).
\end{eqnarray*}
The choice of the number $\zeta$ in definition of the function $\zeta
(y)$ in (\ref{zeta}) guarantees that $B \zeta(y) \leq-C\zeta(y)$.
Moreover, with the earlier assumption that \mbox{$0\leq\sigma< 1-\beta$},
the growth of $\zeta(y)$ as $|y| \rightarrow\infty$ dominates the
growth in $y$ of all other terms in~$H_\varepsilon f_\varepsilon$.
Therefore, there exist constants $c_0, c_1>0$ with
\[
\H f_\ep(x,y)\leq\tfrac12 |\sigma(y) \,\partial_x f(x)|^2 + B \xi(y) -
\theta c_0 \zeta(y) + \ep c_1 .
\]
In addition,
\[
f_\varepsilon(x,y) = f(x) +\varepsilon^3 g(y) \geq f(x) - \varepsilon
^3 \Vert
\xi\Vert_\infty.
\]
Furthermore, for each $c>0$, we can find $K \times\tilde{K} \in
\mathcal Q$, such that
%
%
\begin{equation}
\{ (x,y) \dvtx H_\varepsilon f_\varepsilon(x,y) \geq-c \} \cap\{
(x,y) \dvtx
f_\varepsilon(x,y) \leq c \}\subset K \times\tilde{K}
\end{equation}
verifying Condition \ref{limsupH}\ref{Kf0}. The rest of
Condition \ref
{limsupH} can be verified by taking
\[
H_0(x, p; \xi,\theta) = \sup_{y \in E_0} \biggl( \frac12 |\sigma(y) p|^2
+ B \xi(y) - \theta c_0 \zeta(y) \biggr).
\]
We define
\begin{eqnarray*}
H_0 f(x) :\!&=& \inf_{\alpha\in\Lambda} H_0(x,\partial_x
f(x);\alpha) \\
&=&\inf_{0<\theta<1} \inf_{\xi\in C^2_c(E_0)} \sup_{y \in E_0} \biggl(
\frac12 |\sigma(y) \,\partial_x f(x)|^2 + B \xi(y) - \theta c_0\zeta(y)
\biggr).
\end{eqnarray*}

Similarly, for $f \in D_-$, $\alpha=(\xi, \theta) \in\Lambda$, we
can choose
\[
f_\varepsilon(x,y) = f(x)+ \varepsilon^3 \xi(y) - \varepsilon^3
\theta
\zeta(y).
\]
Then Condition \ref{liminfH} holds for the choice of
\[
H_1(x, p; \xi,\theta) = \inf_{y \in\R} \biggl( \frac12 |\sigma(y)
p|^2 +
B \xi(y) + \theta c_0 \zeta(y) \biggr) .
\]
We define
\begin{eqnarray*}
H_1 f(x) :&\!=& \sup_{\alpha\in\Lambda} H_1(x, \partial_x
f(x); \alpha) \\
&=&\sup_{0<\theta<1} \sup_{\xi\in C^2_c(E_0)} \inf_{y \in E_0} \biggl(
\frac12 |\sigma(y) \,\partial_x f(x)|^2 + B \xi(y) + \theta c_0 \zeta(y)\biggr).
\end{eqnarray*}

Next, to verify Lemma \ref{rigconv}, we estimate $H_0 f$ from above
and $H_1f$ from below using some simple quantity.
%
%
\begin{lemma}
\begin{eqnarray*}
H_0 f(x) &\leq& \tfrac12 |\bar{\sigma} \,\partial_x f(x)|^2,\qquad f \in
D_+;\\
H_1 f(x) &\geq& \tfrac12 |\bar{\sigma} \,\partial_x f(x)|^2,\qquad f \in D_-.
\end{eqnarray*}
We note that $H_0, H_1$ have different domains $D_+$ and $D_-$,
respectively, $D_+ \cap D_- =\varnothing$.
\end{lemma}
\begin{pf} The key to obtaining the estimates in the statement of
the lemma is the Poisson equation,
%
%
\begin{equation}\label{Poissoneqn}
B \chi(y)= \tfrac12 |p|^2 \bigl(\bar{\sigma}^2-\sigma^2(y) \bigr),
\end{equation}
where $B$ is the differential operator (generator of $Y$) defined in
(\ref{B-operator}).
We will need growth estimates for $\chi$. In the case of $\beta=0$
(i.e., $Y$ is an \mbox{O--U} process),
Section 5.2.2 of Fouque, Papanicolaou and Sircar \cite{FPS00} contains
such estimates.
Specifically, if $\sigma(y)$ is bounded, $|\chi(y) | \leq C(1+ \log(1+|y|))$;
if $\sigma(y)$ has polynomial growth, $\chi$ has polynomial growth
estimates of the same order.
The following growth estimates for the situation $\frac1 2\leq\beta
<1$ are derived in Appendix \ref{growthestimates}:
%
%
\begin{equation}\label{boundchi}\qquad
|\chi^\prime(y)|\leq C_1y^{2\sigma-1} \qquad\mbox{as }y\to\infty,
\mbox{for some positive constant }C_1.
\end{equation}
Therefore ${ |\chi(y) | \leq C(1+ \log(1+|y|))}$ if
$\sigma
(y)$ is bounded and
$|\chi(y)|\leq\tilde{C}(1+y^{2\sigma})$ when
$0<\sigma
<1-\beta$.

We will make use of $\chi$ as a test function in the expressions for
$H_0 f$ and $H_1 f$. However, $\chi$ does not have compact support.
We choose a cut-off function $\varphi$ to approximate it using
localization arguments.
Let nonnegative $\varphi(y)\in C^\infty(E_0)$ be such that $\varphi
(y)=1$ when $|y|\leq1$ and $0$ when $|y|>2$. We take a sequence of
$\xi
_n(y) = \varphi(\frac{y}{n}) \chi(y)$, which are truncated versions of
$\chi$. Then
\begin{eqnarray*}
B\xi_n(y)&=&\varphi\biggl(\frac{y}{n} \biggr)B\chi(y)+(m-y)\chi
(y)n^{-1}\varphi
^\prime\biggl(\frac{y}{n} \biggr)\\
&&{} +\frac1 2 \nu^2y^{2\beta}\chi(y)n^{-2}\varphi^{\prime
\prime}
\biggl(\frac{y}{n} \biggr)+\nu^2y^{2\beta}\chi^\prime(y)n^{-1}\varphi
^\prime
\biggl(\frac{y}{n} \biggr).
\end{eqnarray*}
Suppose $\sigma>0$. Noting that $|\varphi(y)|, |\varphi^\prime(y)|$ and
$|\varphi^{\prime\prime}(y)|$ are uniformly bound\-ed and are $0$ when
$|y|>2$, and using the growth estimates (\ref{boundchi}) for $\chi$
and~$\chi^\prime$, we get
\begin{eqnarray*}
|B\xi_n(y)|&\leq& cy^{2\sigma} \biggl(1+\frac{(m-y)}{n}+ \biggl(\frac
{y}{n}
\biggr)^{2\beta}n^{2\beta-2}+y^{\beta-1} \biggl(\frac{y}{n}
\biggr)^\beta n^{\beta-1}
\biggr)1_{\{{y/n}\leq2\}}\\
&\leq& cy^{2\sigma} \qquad\mbox{for all }n.
\end{eqnarray*}
In the above, we used the fact that $\frac y n \leq2$ and $\beta-1<0$.
Similarly, if $\sigma(y)$ is bounded, that is, $\sigma=0$, we get
${ |B\xi_n(y)|}$ is uniformly bounded for all $n$.
Therefore, for large $y$, $\zeta(y)$ dominates $B\xi_n(y)$ uniformly in
$n$ in the following sense: there exists a sub-linear function
$\psi\dvtx\R\mapsto\mathbb R_+$ such that
\[
\sup_{n =1,2,\ldots} |B \xi_n(y)| \leq\psi(\zeta(y)) .
\]

With the above estimate, we have
\begin{eqnarray*}
H_0 f(x) & \leq& \limsup_{n \rightarrow\infty} \inf_{0<\theta<1}
\sup
_{y \in E_0} \biggl( \frac12 |\sigma(y) \,\partial_x f(x)|^2 + B \xi_n(y)
-\theta c_0 \zeta(y) \biggr) \\
& \leq& \frac12 |\bar{\sigma} \,\partial_x f(x)|^2.
\end{eqnarray*}

Similarly, one can prove the case for $H_1f$.
\end{pf}

By standard viscosity solution theory (e.g., \cite{CIL92}), the
comparison principle holds for sub-solutions and super-solutions of
%
%
\begin{eqnarray*}
\partial_t u_0 &=& \tfrac12 |\bar{\sigma} \,\partial_x u_0|^2,\qquad
t >0;
\\
u_0(0,x)&=&h(x),
\end{eqnarray*}
and the solution is uniquely given by the Lax formula (see \cite{Evans97}),
%
%
\begin{equation}\label{lax1}
u_0(t,x) = \sup_{x^\prime\in\R} \biggl\{ h(x^\prime) - \frac
{|x-x^\prime
|^2}{2\bar{\sigma}^2 t} \biggr\}.
\end{equation}
Putting together the above result and Lemmas \ref{sub-super-solns} and
\ref{rigconv}, we get:
%
%
\begin{lemma}
\[
\lim_{\varepsilon\rightarrow0+} \sup_{|t|+|x| + |y| < c}
|u_\varepsilon
(t,x,y) - u_0(t,x)| =0\qquad \forall c>0,
\]
where $u_0$ is the solution of (\ref{u0-H1}) and is given by (\ref{lax1}).
\end{lemma}

\subsection{\texorpdfstring{Case $\delta=\ep^2$}{Case delta = epsilon 2}}\label{delta2-rig}

For each $f =f(x) \in D_+$ and $\alpha=(\xi,\theta) \in\Lambda$, we
choose our perturbed test function as
\[
f_\varepsilon(x,y) := f(x) + \varepsilon g(y),\vadjust{\goodbreak}
\]
where $g(y) = (1-\theta) \xi(y) + \theta\zeta(y)$; $\zeta(y)$ is
defined as before in (\ref{zeta}).
Then
\begin{eqnarray*}
\H f_\ep(x,y)&=&\ep\bigl[\bigl(r-\tfrac12\sigma^2(y)\bigr)\,\partial_x f+\tfrac
12\sigma
^2(y)\,\partial_{xx}^2f \bigr]+\tfrac12\sigma^2(y)|\partial_xf|^2\\
&&{} +e^{-g(y)}B^{\partial_xf(x)}e^{g}(y)\\
&\leq&\ep\bigl[\bigl(r-\tfrac12\sigma^2(y)\bigr)\,\partial_x f+\tfrac12\sigma
^2(y)\,\partial_{xx}^2f \bigr]+\tfrac12\sigma^2(y)|\partial_xf|^2\\
&&{} +(1-\theta) e^{-\xi} B^{\partial_xf} e^\xi(y) + \theta
e^{-\zeta}
B^{\partial_xf} e^\zeta(y),
\end{eqnarray*}
where $B^{\partial_xf(x)}$ is the perturbed generator defined in
(\ref{pertB}).
Recall that\break $\Vert\partial_x f \Vert_\infty+ \Vert\partial^2_{xx} f
\Vert_\infty< \infty$ by the choice of domain $D_+$. We can thus find
a constant $c_0>0$ such that
\[
\H f_\ep(x,y)\leq\tfrac12 |\sigma(y) \partial_x f(x)|^2 + (1-\theta)
e^{-\xi} B^{\partial_xf} e^\xi(y) + \theta e^{-\zeta} B^{\partial_xf}
e^\zeta(y)+ \ep c_0.
\]
Note that
\[
e^{-\zeta} B^{\partial_xf(x)}e^\zeta(y)= B\zeta(y)+\rho\sigma
(y)\nu
y^\beta\,\partial_xf(x)\,\partial_y\zeta(y)+\tfrac1 2 \nu^2y^{2\beta
}|\partial_y\zeta(y)|^2,
\]
where
%
%
\begin{equation}\label{Bzeta}
B\zeta(y)=-\zeta\cdot|y-m|^\zeta+\tfrac12\nu^2y^{2\beta}\zeta
(\zeta
-1)|y-m|^{\zeta-2}.
\end{equation}
The term $- \zeta(y)$ in $B\zeta(y)$ dominates growth in $y$ from all
other terms in $\H f_\ep$ as $|y|\to\infty$. Since $\zeta(y)\to
\infty$
as $|y|\to\infty$, $\H f_{\ep}(x,y)\to-\infty$ as $|y|\to\infty
$. We
also have $f_\varepsilon(x,y) = f(x) +\varepsilon g(y) \geq f(x) -
\varepsilon
\Vert\xi\Vert_\infty$.
Therefore,\vspace*{1pt} for each $c>0$, we can find $K \times\tilde{K} \in
\mathcal
Q$, such that
%
%
\begin{equation}\quad
\{ (x,y) \dvtx H_\varepsilon f_\varepsilon(x,y) \geq-c \} \cap\{
(x,y) \dvtx
f_\varepsilon(x,y) \leq c \}\subset K \times\tilde{K}
\end{equation}
verifying Condition \ref{limsupH}\ref{Kf0}.

The super-solution case follows similarly, where we define the perturbed
test function as ${ f_\ep(x,y)=f(x)+\ep(1+\theta) \xi
(y)-\ep\theta\zeta(y)}$, for each $f\in D_-$ and \mbox{$(\xi,\theta)\in
\Lambda$}.

Take
\begin{eqnarray*}
H_0(x,p; \xi,\theta) &:=& \sup_{y \in E_0} \biggl( \frac12 | \sigma(y)
p|^2 + (1-\theta) e^{-\xi}B^p e^{\xi}(y) +\theta e^{-\zeta} B^p
e^\zeta
(y) \biggr), \\ 
H_1(x,p;\xi,\theta)& :=& \inf_{y \in E_0} \biggl( \frac12 |\sigma(y)
p|^2 + (1+\theta) e^{-\xi} B^pe^{\xi}(y) -\theta e^{-\zeta} B^p
e^\zeta
(y) \biggr)
\end{eqnarray*}
and
\begin{eqnarray*}
H_0 f(x)&: =& \inf_{0<\theta< 1} \inf_{\xi\in C^\infty_c(E_0)}
H_0(x,\partial_x f ; \xi,\theta), \\
H_1f(x) &:=& \sup_{0< \theta<1} \sup_{\xi\in C^\infty_c(E_0)} H_1(x,
\partial_x f; \xi,\theta).
\end{eqnarray*}
Conditions \ref{limsupH} and \ref{liminfH} are satisfied by these
choices of $H_0$ and $H_1$.
Note that, although $\frac12|\sigma(y)p|^2$ is not bounded in $y$, its
growth is at most $|y|^{2\sigma}$ and is dominated by
the growth of $\zeta(y)$ for $|y|$ large enough.\vadjust{\goodbreak}

To verify Lemma \ref{rigconv}, we develop useful sharp estimates for
$H_0$ and $H_1$ next.
Denote
\[
T(t) g(y):= E[ g(Y_t)|Y(0)=y],\qquad g\in C_b(E_0),
\]
and let $\mathbb B$ be the weak infinitesimal generator for semigroup
$\{ T(t)\dvtx t \geq0\}$ in $C_b(E_0)$ (see page 244 of \cite{FK06}
for a
definition of a weak infinitesimal generator). Let
$D^{++}(\mathbb B)$ denote the domain of $\mathbb B$ with functions
strictly bounded from below by a positive constant.
Similarly define notations for $\mathbb B^p$, the weak infinitesimal
generator corresponding to the process $Y^p$ introduced in Section \ref
{heurep2}.
For each $g \in D^{++}(\mathbb B^p)\subset C_b(E_0)$, since $\zeta> 2
\sigma$, there exists compact $K \subset\subset E_0$ with
\begin{eqnarray*}
&&\sup_{y \in E_0} \biggl( \frac12 | \sigma(y) p|^2 + (1-\theta)
\frac
{\mathbb B^p g}{g}(y) +\theta e^{-\zeta} B^p e^\zeta(y)\biggr) \\
&&\qquad = \sup_{y \in K} \biggl( \frac12 | \sigma(y) p|^2 + (1-\theta)
\frac
{\mathbb B^p g}{g}(y) +\theta e^{-\zeta} B^p e^\zeta(y) \biggr).
\end{eqnarray*}
For each $\varepsilon>0$, by truncating and mollifying $g$, we can
find a
$\xi:=\xi_\varepsilon\in C^\infty_c(E_0)$ such that
\[
H_0(x,p;\xi,\theta) \leq\varepsilon+ \sup_{y \in K} \biggl( \frac12 |
\sigma(y) p|^2 + (1-\theta) \frac{\mathbb B^p g}{g}(y) +\theta
e^{-\zeta
} B^p e^\zeta(y) \biggr).
\]

Denote $p=\partial_x f(x)$. Then
%
%
\begin{eqnarray}\label{H0estimate}
H_0 f(x) &\leq&\inf_{0<\theta<1} \inf_{g \in D^{++}(\mathbb B^p)}
\sup
_{y \in E_0} \biggl( \frac12 |\sigma(y) p|^2 + (1-\theta) \frac
{\mathbb
B^p g}{g}(y) \nonumber\\[-8pt]\\[-8pt]
&&\hspace*{156pt}{}+ \theta e^{-\zeta} B^p e^\zeta(y)\biggr).\nonumber
\end{eqnarray}
Similarly, we have
%
%
\begin{eqnarray}\label{H1estimate}
H_1 f(x) &\geq&\sup_{0<\theta<1} \sup_{g \in D^{++}(\mathbb B^p)}
\inf
_{y \in E_0} \biggl( \frac12 |\sigma(y) p|^2 + (1+ \theta) \frac
{\mathbb
B^p g}{g}(y)\nonumber\\[-8pt]\\[-8pt]
&&\hspace*{156pt}{} - \theta e^{-\zeta} B^p e^\zeta(y)\biggr).\nonumber
\end{eqnarray}

We define $I_B(\cdot; p)\dvtx\mathcal P(E_0) \mapsto\R\cup\{
+\infty\}
$ by
\[
I_B(\mu; p): = - \inf_{g \in D^{++}(\mathbb B^p) } \int_{E_0} \frac
{\mathbb B^p g}{g} \,d \mu\wedge\int_{E_0}e^{-\zeta(y)}B^p e^{\zeta
(y)}\,d\mu(y).
\]
However, we can find a sequence $\{g_n\}\subset D^{++}(\mathbb B^p)$
[take, e.g., $g_n: = e^{\zeta_n}$ where $\zeta_n \in C^2_c(E_0)$ are
some smooth truncations of $\zeta$], such that
\[
\int_{E_0}e^{-\zeta(y)}B^p e^{\zeta(y)}\,d\mu(y)\geq\limsup_{n\to
\infty
}\int_{E_0} \frac{\mathbb B^p g_n}{g_n} \,d \mu.
\]
Therefore we have
%
%
\begin{equation}\label{IB}
I_B(\mu; p) = - \inf_{g \in D^{++}(\mathbb B^p) } \int_{E_0} \frac
{\mathbb B^p g}{g} \,d \mu.
\end{equation}

Recall that $Y^p$ denotes the process corresponding to generator $B^p$
(or, equivalently, $\mathbb B^{p}$).
It can be directly verified that $Y^p$ has a unique stationary
distribution $\pi^p$ and that $Y^p$ is reversible with respect to it
(see Appendix \ref{Ypprocess} of this article). Let
\[
\mathcal E^p(f,g) : =- \int f {\mathbb B^p} g \,d \pi^p
\]
be the Dirichlet form for $Y^p$. By the material in Section 7 of
Stroock \cite{Stroock84} (particularly Theorem 7.44; note that the
diffusion generated by $B^p$ has transition density with respect to
Lebesgue measure, e.g., Theorem 4.3.5 of Knight \cite{Knight81}), we get
%
%
\begin{equation}\label{IB-alt}
I_B(\mu;p) = \mathcal E^p\Biggl(\sqrt{\frac{d \mu}{d \pi^p}}, \sqrt
{\frac
{d \mu}{d \pi^p}} \Biggr) = \frac{\nu^2}{2} \int_0^\infty y^{2\beta
}
\Biggl|\partial_y \sqrt{\frac{d \mu}{d \pi^p}}(y)\Biggr|^2
\pi^p(dy);\hspace*{-35pt}
\end{equation}
see Appendix \ref{Adirichletform} for the last equality above.
If $\mu$ in $I_B(\mu;p)$ is not absolutely continuous with respect to
$\pi^p$, then the right-hand quantity in (\ref{IB-alt}) is viewed as
$+\infty$.
Again through Theorem 7.44 of \cite{Stroock84}, we also get that $\bar
{H}_0$, defined in (\ref{DV}), can be expressed as
%
%
\begin{eqnarray}\label{H0var}
\bar{H}_0(p) &=& \sup_{\mu\in\mathcal P(\R_+)} \biggl( \frac{|p|^2}{2}
\int_{\R_+} \sigma^2 \,d \mu- I_B(\mu;p) \biggr) \nonumber\\
&=& \sup_{h \in L^2(\pi^p), \Vert h \Vert_{L^2(\pi^p)} =1} \biggl(
\frac
{|p|^2}{2} \int_{\R_+} \sigma^2(y) h^2(y) \pi^p(dy) \\
&&\hspace*{90.6pt}{} - \frac{\nu^2}{2} \int
_0^\infty y^{2\beta} |\partial h(y)|^2 \pi^p(dy) \biggr).
\nonumber
\end{eqnarray}

As in Lemma 11.35 of \cite{FK06},
\[
\inf_{0<\theta<1} \inf_{g \in D^{++}(\mathbb B^p)} \sup_{y \in E_0}
\biggl( \frac12 |\sigma(y) p|^2 + (1-\theta) \frac{\mathbb B^p
g}{g}(y) +
\theta e^{-\zeta} B^p e^\zeta(y)\biggr)
= \bar{H}_0(p).
\]
Using (\ref{H0estimate}), this immediately gives
\[
H_0 f(x) \leq\bar{H}_0( \partial f(x)),\qquad f \in D_+ .
\]
We will prove a similar inequality estimate for $H_1$, hence giving the
following:

\begin{lemma}\label{estimates}
\begin{eqnarray*}
H_1 f(x) &\geq&\bar{H}_0(\partial f(x)),\qquad f \in D_- ,\\
H_0 f(x)&\leq&\bar{H}_0( \partial f(x)),\qquad f \in D_+ .
\end{eqnarray*}
\end{lemma}

It remains to prove the estimate for $H_1$. By the proof of Lemma B.10
of~\cite{FK06},
%
%
\begin{eqnarray}\label{H1ineq1}
&& \sup_{0<\theta<1} \sup_{g \in D^{++}(\mathbb B^p)} \inf_{y \in\R
_+} \biggl( \frac12 |\sigma(y) p|^2 + (1+ \theta) \frac{\mathbb B^p
g}{g}(y) - \theta e^{-\zeta} B^p e^\zeta(y)\biggr)
\nonumber\hspace*{-35pt}\\[-8pt]\\[-8pt]
&&\qquad \geq\inf_{\nu\in\mathcal P(\R_+), \langle
\zeta, \nu\rangle<+\infty} \liminf_{t \rightarrow\infty} t^{-1}
\log E^{\nu} \bigl[ e^{(1/2) |p|^2 \int_0^t \sigma^2(Y^p_s) \,ds}\bigr].
\nonumber\hspace*{-35pt}
\end{eqnarray}

We show that:
%
%
\begin{lemma}
%
%
\begin{equation}\label{H1ineq2}
\liminf_{t \rightarrow+\infty} t^{-1} \log E\bigl[ e^{(1/2) |p|^2 \int
_0^t \sigma^2(Y^p_s) \,ds }|Y^p_0=y\bigr] \geq\bar{H}_0(p).
\end{equation}
\end{lemma}
\begin{pf}
The proof of (\ref{H1ineq2}) follows essentially the same argument
used in Example B.14 in the Appendix of \cite{FK06}, which we will outline.
Two ingredients need to be emphasized. First, for each $\mu$ with
$I_B(\mu;p)<\infty$, by a~mollification and truncation argument, we can
find a sequence $\mu_n(dy) = \frac{e^{h_n(y)}}{\int e^{h_n}\,d\pi^p} \,d
\pi
^p(y)$ with
$h_n +c_n \in C^\infty_c(E_0)$ for some constant $c_n$, such that
$\lim
_{n \rightarrow\infty}I_B(\mu_n;p) =I_B(\mu;p)$.
Second, for every $y \in E_0$ and every $h \in C^\infty_c(E_0)$, the
following ergodic theorem holds:
%
%
\begin{equation}\label{erg}
\lim_{t \rightarrow\infty} \frac1t E\biggl[ \int_0^t \sigma^2(\tilde
{Y}^h_s)\,ds | \tilde{Y}^h_0=y \biggr] = \int_{-\infty}^\infty\sigma^2(z)
\,d\tilde{\pi}^h(z),
\end{equation}
where
\[
d\tilde{Y}^h_s= \bigl((m-\tilde{Y}^h_s)+\rho p\sigma(\tilde
{Y}^h_s)\nu
(\tilde{Y}^h_s)^\beta+\nu^2 (\tilde{Y}^h_s)^{2\beta}\, \partial
h(\tilde
{Y}^h_s) \bigr)\,ds+\nu(\tilde{Y}^h_s)^\beta\,dW_s^2,
\]
and where $\tilde{\pi}^h$ is the unique stationary distribution of
$\tilde{Y}^h$. We will prove~(\ref{erg}) in Lemma \ref{ergthm}.

The process $\tilde{Y}^h$ is $Y^p$ under the Girsanov transformation
of measures
\[
\frac{dP^h}{dP}\bigg|_{\mathcal F_t}=\exp\biggl\{h(Y_t)-h(Y_0)-\int
_0^te^{-h}B^pe^h(Y_s)\,ds\biggr\},
\]
where $P$ and $P^h$ refer to the probability measures of the processes
$Y^p$ and~$\tilde{Y}^h$, respectively. The invariant distribution of
$\tilde{Y}^h$ is then
\[
d\tilde{\pi}^h=\frac{e^{2h}\,d\pi^p}{\int e^{2h}\,d\pi^p}.
\]
We can write
\begin{eqnarray*}
&&\liminf_{t \rightarrow+\infty} \frac1 t \log E^P \biggl[ \exp
\biggl\{\frac12
|p|^2 \int_0^t \sigma^2(Y^p_s) \,ds \biggr\}\Big|Y^p_0=y \biggr] \\
&&\qquad= \lim_{t\to\infty} \frac1 t \log E^{P^h}\biggl[ \exp\biggl\{
\frac12
|p|^2 \int_0^t \sigma^2(\tilde{Y}^h_s) \,ds\\
&&\qquad\quad\hspace*{92pt}{} - \biggl(h(\tilde
{Y}^h_t)-h(\tilde
{Y}^h_0)\\
&&\qquad\quad\hspace*{111pt}{}-\int_0^te^{-h}B^pe^h(\tilde{Y}^h_s)\,ds
\biggr)\biggr\} \Big\vert\tilde
{Y}^h_0=y\biggr] \\
&&\qquad \geq\lim_{t\to\infty} \frac1 t E^{P^h}\biggl[ \frac12 |p|^2
\int_0^t
\sigma^2(\tilde{Y}^h_s) \,ds\\
&&\qquad\quad\hspace*{54.2pt}{} -\biggl(h(\tilde{Y}^h_t)-h(\tilde
{Y}^h_0)\\
&&\qquad\quad\hspace*{72.3pt}{} -\int_0^te^{-h}B^pe^h(\tilde
{Y}^h_s)\,ds
\biggr)\Big\vert\tilde{Y}^h_0=y\biggr] \\
&&\qquad\quad\mbox{(by Jensen's inequality)}\\
&&\qquad=\frac12 |p|^2 \int_{-\infty}^\infty\sigma^2(z) \,d\tilde{\pi
}^h(z)\\
&&\qquad\quad{}+\int_{-\infty}^\infty e^{-h}B^pe^h(z) \,d\tilde{\pi
}^h(z)\qquad\mbox{(by ergodicity of $\tilde{Y}^h$)}\\
&&\qquad=\frac12 |p|^2 \int_{-\infty}^\infty\sigma^2(z) \,d\tilde{\pi
}^h(z)-\mathcal E^p\Biggl(\sqrt{\frac{d \tilde{\pi}^h}{d \pi^p}},
\sqrt
{\frac{d \tilde{\pi}^h}{d \pi^p}} \Biggr)\\
&&\qquad=\frac12 |p|^2 \int_{-\infty}^\infty\sigma^2(z) \,d\tilde{\pi
}^h(z)-I(\tilde{\pi}^h;p).
\end{eqnarray*}
By arbitrariness of $h$, (\ref{H1ineq2}) follows. To complete the
proof, we finally check that:
%
%
\begin{lemma}\label{ergthm}
Equation (\ref{erg}) holds.
\end{lemma}
\begin{pf}
By It\^o's formula,
\[
E[ \zeta(\tilde{Y}^h_t)] = E[\zeta(\tilde{Y}^h_0)] + E\biggl[\int_0^t
\tilde
{B}^h \zeta(\tilde{Y}^h_s)\biggr]\,ds,
\]
where ${ \tilde{B}^h\zeta(y)= (m-y+\rho p\sigma
(y)\nu
y^\beta+\nu^2 y^{2\beta} \,\partial_y h(y) )\zeta^\prime
(y)+\frac1 2 \nu
^2 y^{2\beta}\zeta^{\prime\prime}(y)}$.
As in (\ref{Bzeta}), $- \zeta(y)$ is the dominating growth term in
$\tilde{B}^h \zeta(y)$. Therefore, defining a family of mean
occupation measure,
\[
\tilde{\pi}^h(t, y, A) := E\biggl[ t^{-1} \int_0^t \mathbf{1}_{\{\tilde{Y}^h_s
\in A\}} \,ds \Big| \tilde{Y}^h_0 =y\biggr],
\]
we have that
\[
\sup_{t >0} \int_z \zeta(z) \tilde{\pi}^h(t,y,dz) = \sup_{t >0} t^{-1}
E\biggl[\int_0^t \zeta(\tilde{Y}^h_s)\,ds \Big| \tilde{Y}^h_0=y \biggr] \leq
C(y;h(\cdot
)) < \infty.
\]
Hence $\{ \tilde{\pi}^h(t,y,\cdot) \dvtx t>0\}$ is tight and along
convergent subsequences and corresponding limiting point $\tilde{\pi
}^h$, we have
%
%
\begin{equation}\label{statsub}
E\biggl[t^{-1} \int_0^t \varphi(\tilde{Y}^h_s)\,ds\Big|\tilde{Y}^h_0=y \biggr]
\rightarrow\int_z \varphi\, d \tilde{\pi}^h,\qquad \varphi\in C_b(E_0).
\end{equation}
Such $\tilde{\pi}^h$ is necessarily a stationary distribution
satisfying $\int\tilde{B}^h \psi d \tilde{\pi}^h=0$ for all $\psi
\in
C^2_c(E_0)$.
Uniqueness of such probability measure can be proved by an argument
similar to the one in Appendix \ref{Ypprocess}. We thus conclude that
there is only one such $\tilde{\pi}^h$ and that convergence
(\ref{statsub}) occurs along the whole sequence, not just subsequences.
Furthermore, the growth of $\sigma^2$ is dominated by
$\zeta$, and so by uniform integrability argument, (\ref{erg}) holds.
\end{pf}
\noqed\end{pf}

Now (\ref{H1estimate}), (\ref{H1ineq1}) and (\ref{H1ineq2}) together
give us the estimate for $H_1$ in Lemma \ref{estimates}.

From (\ref{H0-alt}), we see that $\bar{H}_0(p)$ is convex in $p \in
\R
$. Let us denote its Legendre transform as $\bar{L}_0$, then we have
the following.
%
%
\begin{lemma}
The unique viscosity solution to (\ref{u000}) is
%
%
\begin{equation}\label{lax2}
u_0(t,x) := \sup_{x^\prime\in\R} \biggl\{ h(x^\prime) - t \bar
{L}_0
\biggl(\frac{x-x^\prime}{t}\biggr) \biggr\}.
\end{equation}
Moreover, $\u$ converges uniformly over compact sets in $[0,T]\times
\R
\times E_0$ to~$u_0$.
\end{lemma}
\begin{pf}
We know that $u_0$, defined by (\ref{lax2}), solves (\ref{u000}) by
the Lax formula. That $u_0$ is the unique solution follows from
standard viscosity comparison principle with convex Hamiltonians.
The convergence result follows from multi-scale viscosity convergence
results developed in Section \ref{resultsFK}, Lemmas \ref
{sub-super-solns} and \ref{rigconv}.
\end{pf}

\section{Large deviation, asymptotic for option prices and implied
volatilities}\label{completion}
We finish the proof of Theorem \ref{LDP}, Corollary \ref{option-price}
and Theorem \ref{impliedvol}.

\subsection{A large deviation theorem}

\mbox{}

\begin{pf*}{Proof of Theorem \ref{LDP}} From the previous section we
have $\u(t,x,y)\to u_0(t,x)$ as $\ep\to0$ for each fixed $(t,x,y)\in
[0,T]\times\R\times E_0$.
All we need is exponential tightness of $\{X_{\ep,\delta,t}\}$ to
apply Bryc's lemma and to conclude our proof.
This is obtained as follows.

Let $f(x)=\log(1+x^2)$ and $\zeta(y)$ be defined as in (\ref{zeta}). Take
\[
f_\ep(x,y)=
\cases{f(x)+\ep^3 \zeta(y), &\quad for the case $\delta=\ep^4$,\cr
f(x)+\ep\zeta(y), &\quad for the case $\delta=\ep^2$.}\vadjust{\goodbreak}
\]
Note that $f(x)$ is an increasing function of $|x|$ and $\zeta(\cdot
)\geq0$; therefore, for any $c>0$ there exists a compact set
$K_c\subset\R$ such that $f_\ep(x,y)>c$ when $x\notin K_c$.
We next compute $\H f_\ep(x,y)$ [see (\ref{Hep})]. Observe that since
$\Vert\partial_x f \Vert_\infty+ \Vert\partial^2_{xx} f \Vert
_\infty
< \infty$, by our choice of $\zeta(\cdot)$, $\H f_\ep(x,y)\to
-\infty$
as $|y|\to\infty$.
Therefore $\sup_{x\in R,y\in R}\H f_\ep(x,y)=C<\infty$. For
simplicity, we denote $X_{\ep,\delta,t}$ by $X_{\ep,t}$. The~$P$ and
$E$ below denote probability and expectation conditioned on $(X,Y)$
starting at $(x,y)$.
\begin{eqnarray*}
&&P(X_{\ep,t}\notin K_c)e^{(c-f_\ep(x,y)-tC)/\ep}\\
&&\qquad\leq E \biggl[\exp\biggl\{\frac{f_\ep(X_{\ep,t},Y_{\ep,t})}{\ep
}-\frac{f_\ep
(x,y)}{\ep} \\
&&\qquad\quad\hspace*{34pt}{}-\int_0^te^{-{f_\ep(X_{\ep,s},Y_{\ep,s})}/{\ep
}}A_\ep
e^{{f_\ep(X_{\ep,s},Y_{\ep,s})}/{\ep}}\,ds \biggr\} \biggr]\\
&&\qquad\leq1.
\end{eqnarray*}
In the above inequalities, the term within expectation in the second
line is a~nonnegative local martingale (and hence a~supermartingale);
see \cite{EK85}, Lemma 4.3.2.
We apply the optional sampling theorem to get the last inequality above.
Therefore
\[
\ep\log P(X_{\ep,t}\notin K_c)\leq tC+f_\ep(x,y)-c\leq \mbox{const}-c
\]
giving us exponential tightness of $X_{\ep,t}$.

Let $u_0^{h,r}$ denote the limit of $\ued$ when $\ued(0,x,y)=h(x)$ and
$\delta=\ep^r$, $r=2,4$. Applying Bryc's lemma we get, $\{X_{\ep,\ep
^r,t}\}$ for $r=2,4$ satisfies a LDP with speed $1/\ep$ and rate function
%
%
\begin{equation}\label{LDPcharacterization}I_r(x;x_0,t):= \sup_{h\in
C_b(R)}\{h(x)-u_0^{h,r}(t,x_0)\}.
\end{equation}
In Appendix \ref{ratefn} we check that $I_2(x;x_0,t)=t\bar
{L}_0 (\frac
{x_0-x}{t} )$ where $\bar{L}$ is the Legendre transform of
$\bar{H}_0$
defined in (\ref{DV}), and $I_4=\frac{|x_0-x|^2}{2\bar{\sigma}^2t}$.
\end{pf*}

\subsection{Option prices}

\mbox{}

\begin{pf*}{Proof of Corollary \ref{option-price}}
We follow the proof of Corollary 1.3 in \cite{FFF} and show that
${ \lim_{\ep\to0^+}\ep\log E [ (S_{\ep
,t}-K )^+ ]}$
is bounded above and below by ${ -I_r(\log K;x_0,t)}$.

Recall that we are considering out-of-the-money call options and hence
$x_0<\log K$ [see (\ref{OTM-cond})]. Since our rate functions
$I_r(x;x_0,t)$, for both $r=2,4$, are nonnegative, convex functions
with $I_r(x_0;x_0,t)=0$, they are consequently monotonically increasing
functions of $x$ when $x\geq x_0$. Using this fact and the continuity
of the rate functions, the proof of the lower bound follows verbatim
from the proof in~\cite{FFF}. We refer the reader to \cite{FFF} for details.

The upper bound follows from \cite{FFF} once we justify the following
limit: for any $p>1$,
%
%
\begin{equation}\label{mgf-limit}
\lim_{\ep\to0^+}\ep\log E[S_{\ep,\d,t}^p]=0\qquad \mbox{for both
}\d
=\ep^4\mbox{ and } \d=\ep^2.
\end{equation}
Recall the operator $\Aed$ defined at the beginning of Section
\ref{prelim}. By a slight abuse of notation, we can use $\Aed$ to denote
the operator acting on the unbounded function $e^{px}$ given below:
\[
\Aed e^{px} = \varepsilon\bigl( \bigl( r - \tfrac12 \sigma^2(y)\bigr) p e^{px} +
\tfrac
12 \sigma^2(y) p^2e^{px}\bigr).
\]
Let
\[
M_t:=\exp\biggl\{p\Xedt-pX_{\ep,\d,0}-\int_0^te^{-p\Xeds}\Aed
e^{p\Xeds
}\,ds \biggr\}.
\]
Then $M_t$ is a nonnegative local martingale (supermartingale); this
follows from the proof of \cite{EK85}, Lemma 4.3.2. By the optional
sampling theorem,
\[
EM_t\leq1.
\]
Recall that $\Xedt=\log S_{\ep,\d,t}$, then
%
%
\begin{eqnarray}\label{momentbd}
E[S_{\ep,\d,t}^{p/2}]&=&E[e^{p/2 \Xedt}]\nonumber\\
&\leq&(EM_t)^{1/2} \biggl(E \biggl[\exp\biggl\{pX_{\ep,\d,0}+\int
_0^te^{-p\Xeds
}\Aed e^{p\Xeds}\,ds \biggr\} \biggr] \biggr)^{1/2}\nonumber\\[-8pt]\\[-8pt]
&&\mbox{(by H\"
older's inequality)}\nonumber\\
&\leq&1\cdot e^{px_0/2} \biggl(E \biggl[\exp\biggl\{\int
_0^te^{-p\Xeds}\Aed
e^{p\Xeds}\,ds \biggr\} \biggr] \biggr)^{1/2}.
\nonumber
\end{eqnarray}
We simplify and bound the right-hand side of the above inequality:
\begin{eqnarray*}
&&E\biggl[\exp \biggl\{\int_0^te^{-p\Xeds}\Aed e^{p\Xeds}\,ds
\biggr\}\biggr]\\
&&\qquad=E \biggl[\exp\biggl\{\int_0^t \varepsilon\biggl( \biggl( r - \frac12
\sigma^2(\Yeds)\biggr) p
+ \frac12 \sigma^2(\Yeds) p^2\biggr)\,ds \biggr\} \biggr]\\
&&\qquad=e^{\ep rp t}E \biggl[\exp\biggl\{\d(p^2-p)\int_0^{\ep t/\d}\sigma
^2(Y_{\ep,\d
,(\d/\ep) u})\,du\biggr\} \biggr]\\
&&\qquad\quad\biggl(\mbox{by change of variable }u=\dfrac
\ep\d
s\mbox{; recall that }\d=\ep^2\mbox{ or }\ep^4\biggr)\\
&&\qquad=e^{\ep rp t}E \biggl[\exp\biggl\{\d(p^2-p)\int_0^{\ep t/\d}\sigma
^2(Y_u)\,du\biggr\} \biggr],
\end{eqnarray*}
where $Y_u$ is the process with generator $B$ given in (\ref{B-operator}).
By convexity of exponential functions we get
%
%
\begin{eqnarray}\label{ineq}
&&E \biggl[\exp\biggl\{\int_0^te^{-p\Xeds}\Aed e^{p\Xeds}\,ds
\biggr\} \biggr]\nonumber\\[-8pt]\\[-8pt]
&&\qquad\leq e^{\ep rp t}E \biggl[\frac{\d}{t\ep}\int_0^{\ep t/\d}\exp\{
t\ep
(p^2-p)\sigma^2(Y_u)\}\,du \biggr].
\nonumber
\end{eqnarray}
Since $\d=\ep^2$ or $\ep^4$, $\ep/\d\to\infty$ as $\ep\to0$.
Therefore, by the ergodicity of $Y$ and ${ \exp\{
t(p^2-p)\sigma^2(y)\}\in L^1(d\pi)}$ [this follows from an argument
similar to proof of Lemma \ref{ergthm}; note that $\sigma<1-\beta$ by
Assumption \ref{assmpn1}\ref{Assmsigma}], the right-hand side of the
above inequality (\ref{ineq}) is uniformly bounded for all $\ep>0$.
Putting this together with (\ref{momentbd}), we get (\ref{mgf-limit}).
\end{pf*}

\subsection{Implied volatilities}\label{impliedvolatility}

\mbox{}

\begin{pf*}{Proof of Theorem \ref{impliedvol}}
Recall that $X_{\ep,t}=\log S_{\ep,t}$ and $x_0=\log S_0$.
Note that we have dropped the subscript $\delta$ in the notation and
the dependence on $\delta=\ep^4$ or $\ep^2$ should be understood by context.
Our first step is to show that
%
%
\begin{equation}\label{impliedvollimit}
\lim_{\ep\to0^+}\sigma_\ep(t, \log K, x_0)\sqrt{\ep t}=0.
\end{equation}
Once we have shown this, the rest of the proof is identical to that of
Corollary~1.4 in~\cite{FFF}.

By the definition of implied volatility,
%
%
\begin{eqnarray}\label{impvoldefn}
E[(S_{\ep,t}-K)^+]&=&e^{r\ep t}S_0\Phi\biggl(\frac{x_0-\log K+r\ep
t+
\sigma_\ep^2\ep t/2}{\sigma_\ep\sqrt{\ep t}}
\biggr)\nonumber\\[-8pt]\\[-8pt]
&&{}-K\Phi\biggl(\frac{x_0-\log K+r\ep t- \sigma_\ep
^2\ep t/2}{\sigma
_\ep\sqrt{\ep t}} \biggr),
\nonumber
\end{eqnarray}
where $\Phi$ is the Gaussian cumulative distribution function.
Let $l\geq0$ be the limit of $\sigma_\ep\sqrt{\ep t}$ along a
converging subsequence. If $\lim_{\ep\to0^+}$ of the left-hand side of
(\ref{impvoldefn}) is $0$, then $l$ satisfies
\[
S_0\Phi\biggl(\frac{x_0-\log K}{l}+\frac l 2 \biggr)-K\Phi
\biggl(\frac{x_0-\log K}{
l}-\frac l 2 \biggr)=0.
\]
The only solution of the above equation is $l=0$, and thus we get
(\ref{impliedvollimit}).

We therefore need to prove
%
%
\begin{equation}\label{limit1}
\lim_{\ep\to0^+}E [ (S_{\ep,t}-K )^+ ]=0.
\end{equation}
By (\ref{X}) we have
\[
S_{\ep,t}-K=S_0-K+\ep\int_0^t rS_{\ep,t}\,dt+\sqrt{\ep}\int
_0^tS_{\ep
,t}\sigma(Y_{\ep,t})\,dW^{(1)}_t.\vadjust{\goodbreak}
\]
It can be verified that $E[(S_{\ep,t}-K)-(S_0-K)]^2\to0 $, as $\ep
\to
0$, for both cases $\d=\ep^4$ and $\d=\ep^2$.
Therefore
\[
\lim_{\ep\to0^+}E[(S_{\ep,t}-K)^+]= E[(S_0-K)^+]=0
\]
as $S_0<K$ (this is an out-of-the-money call option).

The same formula is obtained when $S_0>K$ by considering
out-of-the-money put options. We finally turn our attention to
at-the-money implied volatility.
The asymptotic limit of at-the-money (ATM) volatility can be shown to
be $\bar{\sigma}^2$, that is,
\[
\lim_{\ep\to0}\sigma^2_{r,\ep}(t,\log K, x_0)=\bar{\sigma
}^2 \qquad\mbox
{when }x_0=\log K; r=2, 4,
\]
by a similar argument as in \cite{FFF}, Lemma 2.6.
The continuity, at-the-money, of the limiting implied volatility, that
is,
\[
\lim_{|{\log K}-x_0|\to0}\frac{ (\log K- x_0
)^2}{2I_r(\log K,
x_0,t)t}=\bar{\sigma}^2
\]
is obvious in the $r=4$ regime, but is more involved in the $r=2$ regime.
We conjecture that it is true, that is,
%
%
\begin{equation}\label{ATMlimit}
\lim_{z\to0}\frac{z^2}{2t^2\bar{L}_0 ({z/t}
)}=\bar{\sigma}^2,
\end{equation}
and we briefly indicate an outline of the proof.
Let
\[
\Lambda_T(p):=T^{-1} \log E \bigl[ e^{\int_0^T\rho p \sigma(Y_s) \,d
W^{(2)}_s + ({(1-\rho^2)}/{2})|p|^2 \int_0^T \sigma^2(Y_s)
\,ds} \bigr],
\]
so that $\bar{H}_0(p)=\lim_{T\to\infty}\Lambda(p)$. The result
(\ref{ATMlimit}) follows
if $\bar{H}_0(p)$ is twice differentiable in a neighborhood of $p=0$
and $H_0^{\prime\prime}(0)=\frac{\bar{\sigma}^2}{2}$. It can
easily be
checked that $\lim_{T\to\infty}\Lambda_T^{\prime\prime}(0)=\frac
{\bar
{\sigma}^2}{2}$. The main\vspace*{1pt} difficulty is to get a uniform bound on
$\Lambda_T^{\prime\prime\prime}(p)$ for all $T$ and in a neighborhood
of $p=0$.
Obtaining such a uniform bound on $\L'''_T(p)$ involves tedious
calculations but should follow from the multiplicative ergodic
properties of the $Y$ process (see \cite{KM05}).
\end{pf*}

\begin{appendix}
In the following Appendix, we collect some material regarding 1-D
diffusions $Y$ and technical but elementary estimates.

\section{Positivity of the $Y$ process}\label{secFellerclassification}
In this section we prove positivity of the $Y$ process when $\frac1
2<\beta<1$ in~(\ref{Y1}). Assume $m>0$ and $Y_0>0$. Recall the scale
function $s(y)$ defined in the \hyperref[intro]{Introduction}, and let
$S(y)=\int
_1^ys(y)\,dy$. By Lemma 6.1(ii) in Karlin and Taylor \cite{KT81}, to
prove that $Y_t$ remains positive a.s. for all $t\geq0$, it is
sufficient to show that
\[
\lim_{\ep\to0^+}S(\ep)=-\infty.
\]
For $0<\ep\ll1$,
\begin{eqnarray*}
-S(\ep)&=&\int_\ep^1 s(y)\,dy=\int_\ep^1 \exp\biggl\{-\int
_1^y\frac
{2(m-z)}{\nu^2|z|^{2\beta}}\,dz \biggr\}\,dy\\
&=&C\int_\ep^1\exp\biggl\{\frac{2m}{\nu^2(2\beta-1)y^{2\beta
-1}}+\frac
{y^{2-2\beta}}{\nu^2(1-\beta)} \biggr\}\,dy\\
&&\mbox{(where $C$ is a
positive constant and $2\beta-1,1-\beta>0$)}\\
&=&\int_{2\ep}^1(\mbox{positive integrand})\,dy\\
&&{}+ C\int_\ep^{2\ep
}\exp
\biggl\{\frac
{2m}{\nu^2(2\beta-1)y^{2\beta-1}}+\frac{y^{2-2\beta}}{\nu
^2(1-\beta)}
\biggr\}\,dy\\
&\geq& C\ep\exp\biggl\{\frac{2m}{\nu^2(2\beta-1)(2\ep)^{2\beta
-1}} \biggr\}\to
+\infty
\end{eqnarray*}
as $\ep\to0^+$, provided $m>0$.
Therefore $\lim_{\ep\to0^+}S(\ep)=-\infty$.

\section{Growth estimates for solutions to Poisson~equations}
\label{growthestimates}

Assume $\chi$ satisfies the Poisson equation
\[
B \chi(y)= \tfrac12 |p|^2 \bigl(\bar{\sigma}^2-\sigma^2(y) \bigr),
\]
where $\bar{\sigma}^2$, defined in (\ref{sigmabar}), is the
average of $\sigma^2(y)$ with respect to the invariant distribution
$\pi
(dy)$, given in (\ref{pi}), of the $Y$ process. In this section we find
growth estimates for $\chi$.

The right-hand side of the above Poisson equation is centered with
respect to the invariant distribution $\pi(dy)=\frac{m(y)}{Z}\,dy$ [given
in (\ref{pi})], and so
%
%
\begin{equation}\label{centering}
\int_0^\infty m(z)\bigl(\bar{\sigma}^2-\sigma^2(z) \bigr) \,dz=0,
\end{equation}
where
\[
m(y) = \frac{1}{\nu^2 y^{2 \beta}} \exp\biggl\{ \int_1^y \frac{2
(m-z)}{\nu
^2 z^{2 \beta}} \,dz \biggr\}.
\]

By (\ref{decomp}),
\begin{eqnarray*}
\chi(y) :\!&=& \int d S(y) \int_0^y |p|^2 \bigl(\bar{\sigma}^2 - \sigma^2(z)\bigr)
\,d M(z)\\
&=& \int\frac{1}{y^{2 \beta} m (y)} \biggl[\int^y \frac{|p|^2
m(z)(\bar
{\sigma}^2-\sigma^2(z) )}{\nu^2} \,dz \biggr]\,dy
\end{eqnarray*}
is a solution up to a constant, and so
\begin{eqnarray*}
\chi^\prime(y)&=&\frac{|p|^2}{\nu^2y^{2\beta}m(y)} \biggl[\int_0^y
m(z)\bigl(\bar
{\sigma}^2-\sigma^2(z) \bigr) \,dz \biggr]\\
&=&- \frac{|p|^2}{\nu^2y^{2\beta}m(y)} \biggl[\int_y^\infty
m(z)\bigl(\bar{\sigma
}^2-\sigma^2(z) \bigr) \,dz \biggr].
\end{eqnarray*}
The last equality is by the centering condition (\ref{centering}).
Given the bounds on~$\sigma(y)$ in Assumption
\ref{assmpn1}\ref{Assmsigma}, we can compute the following bounds
where the constants,
denoted by $c$, are positive and vary from line to line:
\begin{eqnarray*}
|\chi^\prime(y)|&\leq&\frac{c|p|^2}{\nu^2y^{2\beta}m(y)}\int
_y^\infty
z^{2\sigma}m(z)\,dz\\
&=& \frac{c|p|^2e^{\alpha y^{1-2\beta}}}{\nu^2e^{-({y^{2-2\beta
}})/({\nu
^2(1-\beta)})}}\int_y^\infty z^{2\sigma-2\beta}e^{-\alpha
z^{1-2\beta
}}e^{-({z^{2-2\beta}})/({\nu^2(1-\beta)})}\,dz,
\end{eqnarray*}
where $\alpha=\frac{2m}{\nu^2(2\beta-1)}>0$.
Bounding $e^{-\alpha z^{1-2\beta}}$ above by $1$ we get
\begin{eqnarray*}
|\chi^\prime(y)|&\leq&{ \frac{c|p|^2e^{\alpha
y^{1-2\beta
}}}{\nu^2e^{-({y^{2-2\beta}})/({\nu^2(1-\beta)})}}}\int_y^\infty
z^{2\sigma-2\beta}e^{-({z^{2-2\beta}})/({\nu^2(1-\beta)})}\,dz\\
&=& { \frac{c|p|^2e^{\alpha y^{1-2\beta}}}{\nu
^2e^{-
({y^{2-2\beta}})/({\nu^2(1-\beta)})}}\int_{y^{2-2\beta}}^\infty
u^{
({2\sigma-1})/({2-2\beta})}\exp\biggl\{-\frac{u}{\nu^2(1-\beta
)} \biggr\}\,du}\\
&&\mbox{(by change of variable $u=z^{2-2\beta}$) }\\
&\leq&\frac{c|p|^2e^{\alpha y^{1-2\beta}}}{\nu^2e^{-
({y^{2-2\beta
}})/({\nu^2(1-\beta)})}} \biggl[y^{2\sigma-1}\exp\biggl\{-\frac
{y^{2-2\beta}}{\nu
^2(1-\beta)} \biggr\} \biggr].
\end{eqnarray*}
In the last inequality we used
$\int_{a}^\infty[u^{({2\sigma-1})/({2-2\beta
})}e^{-{u}/({\nu^2(1-\beta)})} ]\,du\leq\nu^2(1-\beta
)a^{({2\sigma
-1})/({2-2\beta})}e^{-{a}/({\nu^2(1-\beta)})}$ (since $\frac
{2\sigma
-1}{2-2\beta}<0$).
Therefore
\[
|\chi^\prime(y)|\leq\frac{c|p|^2e^{\alpha y^{1-2\beta}}}{\nu
^2}y^{2\sigma-1} \sim c|p|^2y^{2\sigma-1} \qquad\mbox{as }y\to
\infty,
\]
since $e^{\alpha y^{1-2\beta}}\sim O(1)$ as $y\to\infty$.

\section{$Y^p$ process}\label{Ypprocess}

Fix $p\in\R$. Denote $\mu_p(y):=(m-y)+\rho p\sigma(y)\nu y^\beta$, and
let $Y^p$ be the process with generator
\[
B^pg=\mu_p(y)\,\partial_yg+\tfrac1 2 \nu^2y^{2\beta}\,\partial
_{yy}^2g,\qquad g\in C^2_c(E_0).
\]
In this section we calculate the unique stationary distribution and
Dirichlet form of the process $Y^p$, and we show that it is a
reversible process. To this end, we first compute the scale function
and speed measure.

The scale function and speed measure for the $Y^p$ process are given by
\[
s_p(y)=\exp\biggl\{-\int_1^y\frac{2\mu_p(z)}{\nu^2z^{2\beta
}} \biggr\} \quad\mbox
{and}\quad m_p(y)=\frac{2}{\nu^2y^{2\beta}s_p(y)}.
\]
Evaluating the integral in $s_p(y)$ we get (the $C$ below denotes a
positive constant that varies from line to line)
\[
s_p(y)=
\cases{
\displaystyle C\exp\biggl\{-\frac{2m\log y}{\nu^2}+\frac{y^{2-2\beta}}{\nu
^2(1-\beta)}
-\frac{2\rho p}{\nu}J \biggr\}, \qquad \mbox{if $\displaystyle \beta=\frac1 2$}, \vspace*{2pt}\cr
\displaystyle C\exp\biggl\{\frac{2m}{\nu^2(2\beta-1)y^{2\beta-1}}+\frac
{y^{2-2\beta}}{\nu
^2(1-\beta)} -\frac{2\rho p}{\nu}J \biggr\}, \vspace*{2pt}\cr
\hphantom{\displaystyle C\exp\biggl\{-\frac{2m\log y}{\nu^2}+\frac{y^{2-2\beta}}{\nu
^2(1-\beta)}
-\frac{2\rho p}{\nu}J \biggr\},\qquad}\hspace*{1.9pt}
\mbox{if $\beta\in
{0}\cup
\biggl(\displaystyle \frac1 2,1\biggr)$},}
\]
where
\[
J(y)=\int^y\frac{\sigma(z)}{z^\beta}\,dz.
\]
Due to bounds on $\sigma$ given in Assumption
\ref{assmpn1}\ref{Assmsigma}, there exist $C_1,C_2>0$ such that
\[
C_1y^{1-\beta}\leq J(y)\leq C_2y^{1-\beta+\sigma},
\]
where
\[
\cases{
0<1-\beta\leq1-\beta+\sigma\leq1, &\quad if $\frac1
2\leq\beta<1$,\vspace*{2pt}\cr
1=1-\beta\leq1-\beta+\sigma<2, &\quad if $\beta=0$.}
\]
Therefore
%
%
\begin{equation}\label{scalefnbounds}
\cases{
\displaystyle \frac{1}{s_p(y)}\to0 \mbox{ when } y\to0 \mbox{ or } y\to\infty,&\quad
if $\displaystyle \frac1 2\leq\beta<1$,\vspace*{2pt}\cr
\displaystyle \frac{1}{s_p(y)}\to0 \mbox{ when } |y|\to\infty, &\quad if $\beta=0$.}
\end{equation}
Define for $y\in E_0$,
\[
S_p(y):=\int_1^ys_p(z)\,dz.
\]
%
Observe that $S_p(y)\to-\infty$ as $y$ approaches the left endpoint of
$E_0$ and $S_p(y)\to+\infty$ as $y\to\infty$.

\subsection{Stationary distribution}\label{Astationarydistn}

Let $\pi^p$ be an invariant distribution of the process $Y^p$. Suppose
it has density function $\Psi(y)$, that is, $d\pi^p(y)=\Psi(y)\,dy$, then
$\Psi$ is uniquely determined as the solution of
\[
\frac1 2 \,\frac{\partial^2}{\partial y^2} (\nu^2y^{2\beta
}\Psi(y)
)-\frac{\partial}{\partial y} (\mu_p(y)\Psi(y) )=0
\]
satisfying $\Psi(y)\geq0$ for all $y$ and $\int_{E_0} \Psi(y)\,dy=1$.
Solving the above differential\vadjust{\goodbreak} equation, we get $\Psi(y)=m_p(y)
[C_1S_p(y)+C_2 ]$.
Since $\Psi$ is nonnegative, and $S_p(y)\to-\infty$ as $y$ approaches
the left boundary of $E_0$, we take $C_1=0$. The other constant $C_2$
is uniquely determined by the condition $\int_{E_0}\Psi(y)\,dy=1$.
Therefore $\pi^p$ is the unique invariant distribution of $Y^p$ and is
given by
%
%
\begin{equation}\label{invdensityYp}\qquad
d\pi^p(y)=\Psi(y)\,dy=\frac{m_p(y)}{Z_1}\,dy=\frac{2}{Z_1\nu
^2y^{2\beta
}s_p(y)}\,dy \qquad\mbox{for } y\in E_0,
\end{equation}
where $Z_1=\int_{E_0} m_p(y)\,dy$.

\subsection{Reversibility} \label{AReversibility}

Let $\varphi,\psi\in C^2_c(E_0)$, then
\begin{eqnarray*}
\int_{E_0} \psi B^p\varphi \,d \pi^p&=&\frac{1}{Z_1} \int_{E_0}\psi
\biggl[\frac1 2 \nu^2y^{2\beta}\varphi^{\prime\prime}+\mu_p(y)\varphi
^\prime
\biggr]\frac{2}{\nu^2y^{2\beta}s_p(y)}\,dy\\
&=&\frac{1}{Z_1}\int_{E_0}\psi\biggl[\varphi^{\prime\prime}
e^{\int^y
{2\mu_p(y)}/({\nu^2y^{2\beta}})}+ \frac{2\mu_p}{\nu^2y^{2\beta
}}\varphi
^\prime e^{\int^y{2\mu_p(y)}/({\nu^2y^{2\beta}})} \biggr]\,dy\\
&=&\frac{1}{Z_1}\int_{E_0}\psi\,\frac{d}{dy} \biggl( \frac{\varphi
^\prime
}{s_p(y)} \biggr)\,dy.
\end{eqnarray*}
Integrating by parts twice and using the boundary conditions (\ref
{scalefnbounds}), we get
\[
\int_{E_0} \psi B^p\varphi\,d \pi^p =\frac{1}{Z_1}\int
_{E_0}\varphi
\,\frac
{d}{dy} \biggl( \frac{\psi^\prime}{s_p(y)} \biggr)\,dy
=\int_{E_0} \varphi B^p\psi\,d \pi^p.
\]

\subsection{Dirichlet form}\label{Adirichletform}

By similar calculations as before, when proving reversibility, we get,
for $f,g\in L^2(\pi^p)$,
\begin{eqnarray*}
\mathcal E^p(f,g):\!&=&- \int_{E_0} f B^pg \,d \pi^p \\
&=& -\frac{1}{Z_1}\int_{E_0} f(y)\,\frac{d}{dy} \biggl(\frac{g^\prime
(y)}{s_p(y)} \biggr)\,dy\\
&=&\frac{1}{Z_1}\int_{E_0} f^\prime(y)g^\prime(y)\,\frac
{1}{s_p(y)}\,dy\\
&=&\frac{\nu^2}{2}\int_{E_0} y^{2\beta} f^\prime(y) g^\prime(y)
\,d\pi^p(y),
\end{eqnarray*}
where we integrated by parts once and used (\ref{scalefnbounds}) in
the second last line.

\section{Rate function formulas}\label{ratefn}

Recall the following characterization of the rate functions given in
(\ref{LDPcharacterization}):
\[
I_r(x;x_0,t)=\sup_{h\in C_b(\R)}\{h(x)-u_0^{h,r}(t,x_0)\},
\]
where
$r=2,4$ correspond to the two regimes $\d=\ep^2$ and $\d=\ep^4$,
respectively.
The $u_0^{h,r}$ are given in (\ref{lax2}) and (\ref{lax1}),
respectively, as
\begin{eqnarray*}
u^{h,2}_0(t,x_0)&=&\sup_{x^\prime\in\R} \biggl\{h(x^\prime)-t\bar
{L} \biggl(\frac
{x_0-x^\prime}{t} \biggr) \biggr\},\\
u^{h,4}_0(t,x_0)&=&\sup_{x^\prime\in\R} \biggl\{h(x^\prime)-
\biggl(\frac
{|x_0-x^\prime|^2}{2\bar{\sigma}^2t} \biggr) \biggr\}.
\end{eqnarray*}

For notational convenience, we will drop the subscript $r$ in $I_r$,
and, in the case $r=4$, we will denote the term $ (\frac
{|x_0-x^\prime
|^2}{2\bar{\sigma}^2t} )$ by $t\bar{L} (\frac
{x_0-x^\prime}{t} )$.
The rate functions can then be rewritten as
\[
I(x;x_0,t)=\sup_{h\in C_b(\R)}\inf_{x^\prime\in\R} \biggl\{
h(x)-h(x^\prime
)+t\bar{L} \biggl(\frac{x_0-x^\prime}{t} \biggr) \biggr\}
\]
for both regimes $r=2$ and $r=4$.
%
%
\begin{lemma}\label{ratefn-2}
\[
I(x;x_0,t)=t\bar{L} \biggl(\frac{x_0-x}{t} \biggr).
\]
\end{lemma}
\begin{pf}
Note that for both cases $r=2,4$, $\bar{L}_0$ is convex, $\bar
{L}_0(0)=0$ and~$\bar{L}_0$ is a nonnegative function. This is obvious
for the case $r=4$. We can deduce this in the $r=2$ case since $\bar
{H}_0(p)$ [defined in (\ref{DV})] is convex and $\bar{H}_0(0)=0$.

Re-write
\begin{eqnarray*}
I(x;x_0,t)&=&t\bar{L}_0 \biggl(\frac{x_0-x}{t} \biggr)\\
&&{}+\sup_{h\in C_b(\R)}\inf
_{x^\prime\in\R}\biggl\{
h(x)-h(x^\prime)
+t\bar{L}_0 \biggl(\frac{x_0-x^\prime}{t} \biggr)-t\bar{L}_0 \biggl(\frac
{x_0-x}{t} \biggr)\biggr\} \\
&=&t\bar{L}_0 \biggl(\frac{x_0-x}{t} \biggr)+ J,
\end{eqnarray*}
where $J= {\sup_{h\in C_b(\R)} J_h }$ and
$J_h=\inf_{x^\prime\in\R} \{h(x)-h(x^\prime)+t\bar{L}_0 (\frac
{x_0-x^\prime}{t} )-t\times\bar{L}_0 (\frac{x_0-x}{t} )
\}$.
Taking $x^\prime=x$ in the $\inf$ we get $J_h\leq0$
and therefore
%
%
\begin{equation}\label{Jleq2}
J\leq0.
\end{equation}

Note that $x_0$ and $x$ are fixed. Define a function $h^*\in C_b(\R)$
as follows:
\[
h^*(x^\prime)=t\bar{L}_0 \biggl(\frac{x_0-x^\prime}{t} \biggr)\wedge t\bar
{L}_0 \biggl(\frac{x_0-x}{t} \biggr).
\]
Then
\[
J_{h^*}=0
\]
and consequently
%
%
\begin{equation}\label{Jgeq2}
J\geq0.
\end{equation}
By (\ref{Jleq2}) and (\ref{Jgeq2}), $J=0$ and we get
\[
I(x;x_0,t)=t\bar{L}_0 \biggl(\frac{x_0-x}{t} \biggr).
\]
\upqed\end{pf}
\end{appendix}

\section*{Acknowledgments}

The authors are greatly indebted to two anonymous referees, whose
suggestions greatly improved both the content and readability of this
paper.


%

\printaddresses

\end{document}